%% file: paper.tex
\pdfoutput=1
\documentclass[aps,prd,amsmath,floats,floatfix, twocolumn,
superscriptaddress,nofootinbib,showpacs,longbibliography]{revtex4-2}

\usepackage[T1]{fontenc}
\usepackage[utf8]{inputenc}
\usepackage{lmodern}

\usepackage{verbatim}

\usepackage[dvipsnames, usenames]{xcolor}
\definecolor{linkcolor}{rgb}{0.0,0.3,0.5}
\usepackage[hypertexnames=false, unicode, colorlinks=true, linkcolor=linkcolor,
citecolor=linkcolor, filecolor=linkcolor,urlcolor=linkcolor,
pdfusetitle]{hyperref}

\usepackage[all]{hypcap}
\usepackage{graphicx}
\usepackage{xspace}
\usepackage{amssymb}

\usepackage[normalem]{ulem} 
\usepackage{bm} 

\usepackage{microtype}

\usepackage{tikz}
\usetikzlibrary{calc, intersections, arrows.meta, positioning, decorations.text, decorations.pathmorphing}

\usepackage{orcidlink}

\graphicspath{%
  {figs/}%
}

\DeclareMathAlphabet{\mathpzc}{OT1}{pzc}{m}{it}

\newcommand{\onecol}{3.4in}
\newcommand{\twocol}{7.0625in}

\newcommand{\codename}[1]{\texttt{#1}}

\begin{document}

\title{Optimizing post-Newtonian parameters and fixing the BMS frame for numerical-relativity waveform hybridizations}

\newcommand{\Cornell}{\affiliation{Cornell Center for Astrophysics
    and Planetary Science, Cornell University, Ithaca, New York 14853, USA}}
\newcommand\CornellPhys{\affiliation{Department of Physics, Cornell
    University, Ithaca, New York 14853, USA}}
\newcommand\Caltech{\affiliation{Theoretical Astrophysics 350-17, California Institute of
    Technology, 1200 E California Boulevard, Pasadena, CA 91125, USA}}
\newcommand{\AEI}{\affiliation{Max Planck Institute for Gravitational Physics
(Albert Einstein Institute), D-14476 Potsdam, Germany}}
\newcommand{\UMassD}{\affiliation{Department of Mathematics,
    Center for Scientific Computing and Visualization Research,
    University of Massachusetts, Dartmouth, MA 02747, USA}}
\newcommand\UMiss{\affiliation{Department of Physics and Astronomy,
    University of Mississippi, University, MS 38677, USA}}
\newcommand{\Bham}{\affiliation{School of Physics and Astronomy and Institute
    for Gravitational Wave Astronomy, University of Birmingham, Birmingham, B15
    2TT, UK}}

\author{Dongze Sun
\orcidlink{0000-0003-0167-4392}}
\email{dzsun@caltech.edu}
\Caltech
\author{Michael Boyle
\orcidlink{0000-0002-5075-5116}}
\Cornell
\author{Keefe Mitman
\orcidlink{0000-0003-0276-3856}}
\Caltech
\author{\\ Mark A.\ Scheel~\orcidlink{0000-0001-6656-9134}}
\Caltech
\author{Leo C.\ Stein
  \orcidlink{0000-0001-7559-9597}}
\UMiss
\author{Saul A.\ Teukolsky
\orcidlink{0000-0001-9765-4526}}
\Caltech
\Cornell
\author{Vijay Varma
\orcidlink{0000-0002-9994-1761}}
\UMassD
\AEI

\hypersetup{pdfauthor={Sun et al.}}

\date{\today}

\begin{abstract}
Numerical relativity (NR) simulations of binary black holes provide precise waveforms, but are typically too computationally expensive to produce waveforms with enough orbits to cover the whole frequency band of gravitational-wave observatories. Accordingly, it is important to be able to hybridize NR waveforms with analytic, post-Newtonian (PN) waveforms, which are accurate during the early inspiral phase. We show that to build such hybrids, it is crucial to both fix the Bondi-Metzner-Sachs (BMS) frame of the NR waveforms to match that of PN theory, and optimize over the PN parameters to mitigate the error caused by the discrepancy between NR and PN parameters. We test such a hybridization procedure including all spin-weighted spherical harmonic modes with $|m|\leq \ell$ for $\ell\leq 8$, using 29 NR waveforms with mass ratios $q\leq 10$ and spin magnitudes $|\chi_1|, |\chi_2|\leq 0.8$. We find that for spin-aligned systems, the PN and NR waveforms agree very well. The difference is limited by the small nonzero orbital eccentricity of the NR waveforms, or equivalently by the lack of eccentric terms in the PN waveforms. To maintain full accuracy of the simulations, the matching window for spin-aligned systems should be at least 5 orbits long and end at least 15 orbits before merger. For precessing systems, the errors are larger than for spin-aligned cases. The errors are likely limited by the absence of precession-related spin-spin PN terms. Using $10^5\,M$ long NR waveforms, we find that there is no optimal choice of the matching window within this time span, because the hybridization result for precessing cases is always better if using earlier or longer matching windows. We provide the mean orbital frequency of the smallest acceptable matching window as a function of the target error between the PN and NR waveforms and the black hole spins.
\end{abstract}

\maketitle

\section{Introduction}
\label{sec:introduction}
Gravitational-wave (GW) astronomy has revealed previously unattainable details about the astrophysics of compact objects \cite{TheLIGOScientific:2016agk,TheLIGOScientific:2016pea,TheLIGOScientific:2016qqj,TheLIGOScientific:2016wfe,TheLIGOScientific:2017qsa,LIGOScientific:2018jsj,LIGOScientific:2021djp,LIGOScientific:2020stg,LIGOScientific:2021psn} and the behavior of gravity in extreme regimes \cite{TheLIGOScientific:2016src,LIGOScientific:2019fpa,LIGOScientific:2021sio}, thanks to the collaboration of LIGO, Virgo, and KAGRA \cite{TheLIGOScientific:2014jea,LIGOScientific:2018mvr,TheVirgo:2014hva,KAGRA:2018plz}.
To further advance the field, the next generation of ground-based detectors, such as the Einstein Telescope (ET) \cite{Punturo:2010zz} and Cosmic Explorer (CE) \cite{Reitze:2019iox}, as well as the first-generation of space-based detectors, including LISA \cite{Amaro-Seoane:2012aqc}, TianQin \cite{TianQin:2015yph}, and Taiji \cite{Hu:2017mde}, are currently under development.
These detectors are expected to provide a significant increase in sensitivity and bandwidth, enabling the detection of GW signals at frequencies as low as 0.1 mHz \cite{Maggiore:2019uih,Amaro-Seoane:2012vvq,Bailes:2021}.

To fully exploit the scientific potential of these gravitational-wave observations, it is essential to construct precise theoretical waveform templates. 
The most accurate waveform templates are obtained through numerical simulations of the full set of Einstein's equations of general relativity, which is known as numerical relativity (NR). 
However, NR simulations are typically too computationally expensive to produce waveforms with enough orbits to cover the whole frequency band that is detectable by gravitational-wave observatories. 
Alternatively, analytic models based on the post-Newtonian (PN) approximation are efficient at providing waveforms that cover the low-frequency band during the early inspiral phase of binary black hole (BBH) systems. 
However, PN waveforms are not accurate for the late inspiral and merger phases of BBH systems. 
Consequently, it is important to be able to attach a PN waveform to the beginning of an NR waveform, thus producing a \emph{hybrid} waveform that is accurate for
the entire frequency band of GW detectors.

Such hybridization procedures were developed, for example, in Refs.~\cite{Santamaria:2010yb,MacDonald:2011ne,Boyle:2011dy,MacDonald:2012mp,Varma:2018mmi,Sadiq:2020hti}. These procedures attained mismatches of $10^{-4}$ between the hybrid and NR waveforms in the matching window (for $m\neq0$ modes) for non-precessing systems \cite{Varma:2018mmi}, and between $10^{-2}$ and $10^{-1}$ (for $m\neq0$ modes) for precessing systems \cite{Sadiq:2020hti}. However, these procedures have four major limitations:
(1) They did not account for the $m=0$ modes since their NR waveforms did not include the correct gravitational-wave memory contribution.
(2) The NR waveforms used in these procedures were not in the same frame as the PN waveforms.
Reference~\cite{Mitman:2022kwt} showed that comparing CCE waveforms without fixing the gauge freedom can result in relative errors three orders of magnitude larger than if one first fixes such freedoms. 
And they also showed that comparing CCE waveforms after mapping them to the same BMS frame can achieve smaller errors than comparing the extrapolated waveforms.
Furthermore, Ref.~\cite{Mitman:2022kwt} successfully developed a hybrid procedure that incorporates memory-containing waveforms.
(3) These procedures used NR values of masses and spins to compute the corresponding PN waveform.
However, the mass and spin parameters are defined differently in PN than they are in NR, and hence the NR and PN parameters are expected to agree only in the limit of infinite binary separation, even for infinite-resolution NR and infinite-order PN. 
In addition, for finite-resolution NR and finite-order PN there are systematic errors in the mapping from parameters to waveforms.
(4) Existing procedures are primarily applicable to spin-aligned or anti-aligned systems, such as binaries born from isolated stellar evolution \cite{Bavera:2020inc,Ma:2023nrf,Gerosa:2018wbw}.
The accuracy of hybridization for precessing systems is severely limited by the inaccurate methodology previously mentioned: the mismatch is on the order of $10^{-2}$ or higher~\cite{Sadiq:2020hti}.
Theoretical considerations~\cite{Bavera:2020inc,Gerosa:2018wbw} and observational evidence~\cite{LIGOScientific:2021psn,Tiwari:2018qch,Rodriguez:2016vmx} indicate that spins can be randomly oriented for binaries that form dynamically in dense environments, and even binaries formed in isolation can exhibit distinctive precessional dynamics.
Therefore, it is crucial to develop accurate hybridization procedures that are valid for precessing systems.

One technique that enables the extraction of waveforms at null infinity is Cauchy-characteristic evolution (CCE)~\cite{Bishop:1996gt,Winicour:2008vpn,Babiuc:2011qi,Moxon:2020gha,Moxon:2021gbv}.
It uses a worldtube from the Cauchy evolution of the Einstein field equations as the inner boundary to a characteristic evolution along null rays. 
The gravitational information is propagated and evolved to future null infinity, so one can obtain waveforms in any desired gauge. 
Reference~\cite{Mitman:2022kwt}
has shown that by mapping the CCE waveform to the correct Bondi-Metzner-Sachs (BMS) frame, the NR waveform agrees much better with the PN waveform.
A key advantage of the hybrid waveforms presented here is that we use CCE NR waveforms. By contrast, previous hybrid waveform models \cite{Santamaria:2010yb,MacDonald:2011ne,Boyle:2011dy,MacDonald:2012mp,Varma:2018mmi,Sadiq:2020hti} used NR waveforms that were computed at null infinity using an extrapolation technique \cite{Boyle:2009vi} that does not have a well-defined BMS frame and does not yield the correct memory contribution.

\begin{figure*}[t]
    \centering
    \input{figs/mapping.tex}
    \caption{%
      \label{fig:mapping-cartoon}%
      NR and PN parameter spaces are distinct because of the different gauges and different definitions of
parameters used in PN and NR models.  The maps into
      infinite-dimensional waveform space differ at each NR resolution
      and PN order.  We optimize PN parameters and adjust the BMS frame to minimize the
      ``vertical'' difference transverse to the images of NR and PN
      parameter spaces.%
    }
\end{figure*}
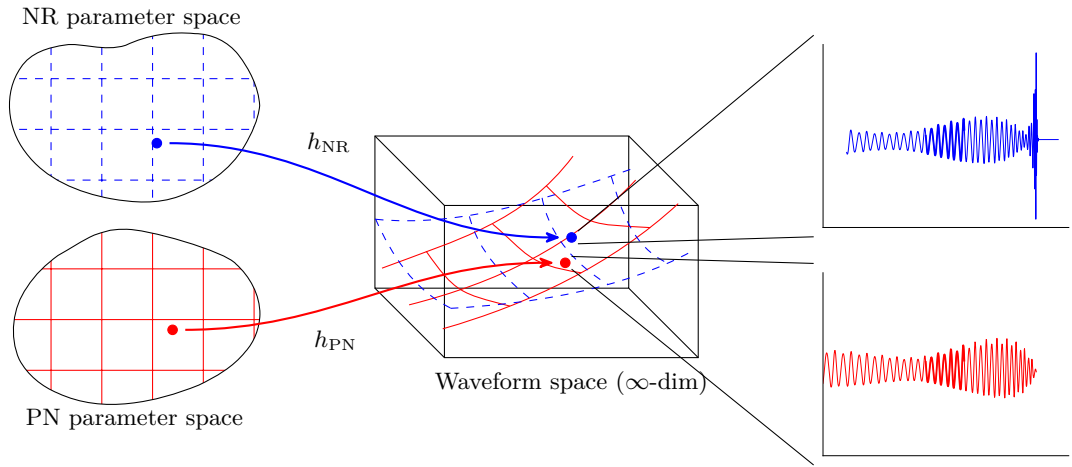

In this work, we simultaneously optimize over the PN parameters and fix the BMS frame of NR waveforms to build hybrid waveforms, so that we can mitigate the error caused by the discrepancy between NR and PN parameters, and inaccurate BMS frame.
This is depicted graphically in Fig.~\ref{fig:mapping-cartoon}.
The optimization and the fixing of the BMS frame, which includes the alignment of the rotation and time shift, are performed inside a time interval called a
\emph{matching window}.
Because we use CCE and thus obtain the correct gravitational-wave memory, we are able to include all $-\ell\leq m \leq \ell$ for $\ell\leq 8$ spin-weighted spherical harmonic modes in the hybrid waveforms.
We test the hybridization procedure using 29 NR waveforms from the SXS catalog~\cite{SXSCatalog} with mass ratios $q\leq 10$ and $|\chi_1|, |\chi_2|\leq 0.8$.
The length of these waveforms ranges from 60 to over 100 orbits, making it possible to test the hybrid waveforms by comparing them to long NR waveforms at a very early stage.
We find that for spin-aligned systems, we can reduce the PN--NR
mismatches to the error caused by small non-zero eccentricities of the NR systems.
We achieve a mismatch (in a 10-orbit long matching window ending
$8000\,M$ before merger) around two to three orders of magnitude
smaller than previous
models~\cite{Santamaria:2010yb,MacDonald:2011ne,Boyle:2011dy,MacDonald:2012mp,Varma:2018mmi}.
For precessing systems, we find that hybridization is likely limited
by the absence of precession-related spin-spin PN terms, yielding
mismatches one to two orders of magnitude smaller than previous
models~\cite{Sadiq:2020hti}.

This paper is organized as follows.
In Sec.~\ref{sec:waveforms}, we introduce the PN and NR waveforms that
we use to build the hybrid waveforms, and we discuss the freedoms in
the physical parameters, 3D rotations and time
shift, and further BMS frame parameters not previously considered.
In Sec.~\ref{sec:PNparameter}, we explain why there are discrepancies between PN and NR parameters, and how we match the physical parameters and inertial frames between NR and PN waveforms.
Following this, we present the complete procedure to build the hybrid waveforms in Sec.~\ref{sec:Procedure}.
In Sec.~\ref{sec:results}, we show the hybridization results as well as the improvement from PN parameters and BMS frame fixing compared with previous models, and we use the findings to more precisely compare the PN and NR waveforms.
Sec.~\ref{sec:error} analyzes the sources of error for spin-aligned and precessing systems.
Additionally, based on our error analysis, we make recommendations for future surrogate models' matching window selection in Sec.~\ref{sec:WindowChoice}.
In Sec.~\ref{sec:conclusion}, we provide a few closing remarks.
We make the code used for this hybridization procedure publicly available in Ref.~\cite{Sun:HybridizationWaveforms}.

\section{Waveforms}
\label{sec:waveforms}
The complex strain $h=h_+ - ih_\times$ is decomposed into modes according to
\begin{equation}
h(u,\theta,\phi) = \sum_{\ell=2}^\infty\sum_{m=-\ell}^\ell h_{\ell m}(u)\ _{-2}Y_{\ell m}(\theta,\phi),
\end{equation}
where $u\equiv t-r$ is the retarded time and $\phantom{}_{-2}Y_{\ell m}$ are the spin-weight $-2$ spherical harmonics. We model all the $-\ell\leq m\leq \ell$ modes for $\ell\leq8$.

\subsection{Post-Newtonian waveforms}\label{sec:PNwaveforms}
In this study, we use PN theory to describe the GW waveforms of BBHs in quasi-circular orbits during the early inspiral stage, and include GW memory contributions.
Memory effects should be inherent in any correct PN formulation, but their importance can often be overlooked. 
In our case, we find that accounting for memory is crucial when comparing to NR waveforms which have memory (since memory effects inherently arise when directly solving Einstein's equations), and thus crucial for accurate hybridization.

For binding energy and flux, we include nonspinning terms that are complete to 3.5 PN order from \cite{Blanchet:2013haa}, and also the 4 PN terms from \cite{Blanchet:2023sbv}.
We also include higher-order terms up to 6 PN from \cite{Fujita:2012cm}, but these terms are obtained in the EMRI limit.
We include the spin-orbit terms up to 4 PN order from \cite{Marsat:2013caa}, spin-spin terms up to 3 PN order from \cite{Bohe:2015ana}, and 3.5 PN spin-cubed terms from~\cite{Marsat:2014xea} for the orbital phase.

For the precession of spins and the orbital angular momentum, we include spin-orbit terms up to 3.5 PN from \cite{Bohe:2012mr}, and spin-spin terms up to 4 PN from \cite{Bohe:2015ana}.

For mode amplitudes, we include terms up to 4 PN. The terms up to 3 PN
are taken from \cite{Blanchet:2008je}, the 3.5 PN terms in the $(2,2)$ mode and $\ell=3$ modes are taken from~\cite{Faye:2012we} and \cite{Faye:2014fra}, respectively, and the 4 PN terms in $(2,2)$ mode are taken from \cite{Blanchet:2023sbv}.
We also include non-linear memory terms up to 3.5 PN \cite{Favata_2009,Mitman:2022kwt} for mode amplitudes, but they are only complete at 3 PN for non-spinning systems, and 2 PN for spinning systems. They are computed using the procedure outlined in Ref.~\cite{Mitman:2022kwt}.

We also generate the PN Moreschi supermomentum~\cite{Moreschi:1988pc}, which we will later use to fix the BMS frame of NR systems, as we will detail in Sec.~\ref{sec:BMS}.
The Moreschi supermomentum modes are taken from \cite{Mitman:2022kwt}, which has the leading order spin-spin and spin-orbit terms.
We leave the effects of eccentric orbits for discussion
in Sec.~\ref{sec:error}. 

To produce the PN phase as a function of time,
we use the TaylorT1 approximant. 
The difference between different PN approximants is anticipated to be at the same order with the PN truncation error, which is controlled by the PN truncation order, implying that switching PN approximants is unlikely to substantially reduce the hybridization error.
The PN generation is implemented in the Python package \codename{NRPNHybridization}~\cite{Sun:HybridizationWaveforms}.

There are 8 independent physical parameters in quasicircular PN theory: mass ratio $q$, total mass $M$, and two three-dimensional spin vectors $\bm{\chi}_1$ and $\bm{\chi}_2$.
The remaining parameters are all gauge.
There are 4 parameters for a time shift and $\mathrm{SO}(3)$ rotation, which
were considered in previous hybridization work.
We will determine the above 12 parameters by performing a 12D optimization.
The rest of the BMS group includes a Lorentz boost (3 parameters),
and a supertranslation (described below), which is given by a smooth
real-valued function $\alpha(\theta,\phi)$ on the sphere.
We band-limit $\alpha$ to $\ell \le 8$ modes, and its (0,0) mode is already
accounted for in the time shift above, so $\alpha$ contributes an additional
$(\ell+1)^{2}-1=80$ real parameters.
We will discuss how to determine these 80 parameters in the next subsection.

\subsection{Numerical-relativity waveforms and BMS frames}
\label{sec:BMS}
It is worth noting that models of gravitational-wave emission are usually constructed in asymptotically flat spacetimes.  
Bondi, van~der~Burg, Metzner, and Sachs~\cite{Bondi:1962px, Sachs:1962wk, Sachs:1962zza} demonstrated that the spacetime symmetry at the asymptotic boundary is the BMS group.  This is comprised of the usual Lorentz transformations, as well as a set of generalized time and space translations referred to as ``supertranslations,'' which can be interpreted as direction-dependent translations on the asymptotic boundary.
Specifically, under a supertranslation $\alpha(\theta,\phi)$, the retarded-time coordinate $u \equiv t-r$ transforms as
\begin{equation}
  u' = u - \alpha(\theta,\phi)
  \,.
\end{equation}
These transformations are implemented in the \texttt{scri} python package~\cite{scri}, using the method described in Ref.~\cite{Boyle:2015nqa}.

Note that the BMS frame is defined at null infinity, but the outer boundary of NR simulations is set at a finite radius, which doesn't have the information from null infinity.
So the NR BMS frame can not be specified arbitrarily when setting up the simulation, and therefore in practice an NR simulation (before any BMS frame fixing is done) is in some arbitrary uncontrolled BMS frame.
Reference~\cite{Mitman:2021xkq} emphasized the importance of having both PN and NR waveforms in the same BMS frame.
We follow their procedure to map NR waveforms to the corresponding PN BMS frame.
First, we determine the space translation and boost by minimizing the
center-of-mass charge. We then calculate the supertranslations that map the $\ell\geq 2$ modes of the NR Moreschi supermomentum $\Psi^{\mathrm{NR}}_{\mathrm{M}}$ to the corresponding PN Moreschi supermomentum $\Psi^{\mathrm{PN}}_{\mathrm{M}}$, where $\Psi_{\mathrm{M}}$ is
\begin{equation}
\Psi_{\mathrm{M}} \equiv \Psi_2+\sigma \dot{\bar{\sigma}}+\eth^2 \bar{\sigma},
\label{eq:SuperMomentum}
\end{equation}
with $\Psi_2$ being the Weyl scalar with spin-weight 0, $\sigma$ the Newman-Penrose shear, and an overbar representing complex conjugation.
For a quantity $\eta$ of spin weight $s$, the spin-weight raising operator $\eth$ can be written as \cite{Newman:1966ub, Boyle:2016tjj}
\begin{equation}
\eth\eta = -\frac{1}{\sqrt{2}}(\sin\theta)^s\left(\frac{\partial}{\partial\theta}+\frac{i}{\sin\theta}\frac{\partial}{\partial\phi}\right)\left((\sin\theta)^{-s}\eta\right).
\end{equation}
The mapping to the BMS frame is done using an iterative procedure~\cite{Mitman:2021xkq, Mitman:2022kwt}, as we explain in more detail below in Sec.~\ref{sec:Procedure}.

We use NR waveforms that are computed using a Cauchy-characteristic evolution (CCE) \cite{Moxon:2020gha,Moxon:2021gbv}, for which the Cauchy evolution parts are performed using the Spectral Einstein Code (SpEC) \cite{SpECwebsite}, and the CCE computation is performed using the SpECTRE code \cite{spectrecode}.
CCE uses a worldtube from the Cauchy evolution of the Einstein field equations as the inner boundary of a characteristic evolution on a null foliation. 
The gravitational information is propagated and evolved to future null infinity. 
Note that $\Psi_2$ and $\sigma$ required in Eq.~\eqref{eq:SuperMomentum} are calculated by the SpECTRE CCE code. Note that unlike CCE waveforms, extrapolated waveforms obtained using the methods described in Ref.~\cite{Boyle:2009vi} cannot be mapped to the PN BMS frame, because they do not have an accurate value of $\Psi_2$ for calculating the Moreschi supermomentum.

\begin{figure}[tb]
\includegraphics[width=\onecol]{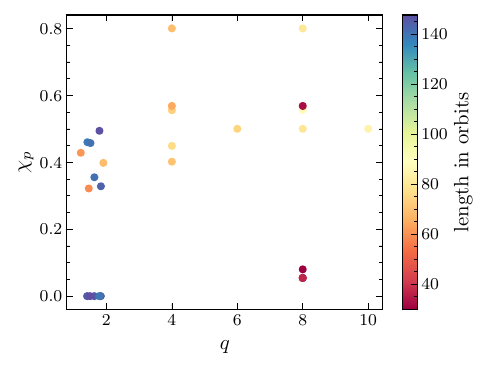}
\caption{The parameter space of the NR simulations that we use. The color shows the length of the waveform in terms of orbits.}
\label{fig:Systems}
\end{figure}

The NR waveforms that we use are from simulations in the SXS catalog \cite{SXSCatalog,Boyle:2019kee}.
We plot the parameter space $\chi_p$ versus $q$ in Fig.~\ref{fig:Systems}, where $\chi_p = \text{max}\{\chi_{1\perp}, \frac{4+3q}{q(4q+3)}\chi_{2\perp}\}$ \cite{Schmidt:2014iyl} and $\chi_{1,2\perp}$ is the magnitude of the projection in the orbital plane.
They are systems SXS:BBH:1412--1416 and 2617--2640.
Of these, SXS:BBH:1412--1416 and 2617-2621 are exceptionally long, spanning over 100 orbits. This extensive length enables us to evaluate the validity of hybridization at an early stage in the binary's inspiral. 
The mass ratio for both the SXS:BBH:1412--1416 and 2617-2635 systems falls within the range $1<q<10$. 
As for the SXS:BBH:2636--2640 systems, the mass ratios are approximately $q\approx8$.
The waveforms SXS:BBH:1412--1416 correspond to spin-aligned systems, while the other systems utilized in our study are precessing systems. 
The magnitudes of the spin parameters $\chi_1$ and $\chi_2$ are constrained as
$|\chi_1|, |\chi_2|\leq 0.8$.

The 8 physical parameters in NR simulations, namely the masses $m_1$ and $m_2$ and dimensionless spin vectors $\bm{\chi}_1$ and $\bm{\chi}_2$, are measured quasi-locally on the apparent horizons of the BHs \cite{Boyle:2019kee}.
Therefore, the value of these quantities under this `quasi-local' definition could differ from the PN mass and spin parameters that appear in PN expansions. 
This difference is accounted for in the matching procedure.


\section{Matching PN and NR parameters}
\label{sec:PNparameter}

There are three reasons why we want to perform the optimization over all intrinsic parameters:

\begin{enumerate}
\item The definitions of PN parameters are not consistent with the NR
  definitions.  The NR parameters are measured quasi-locally on the
  apparent horizon, as detailed in~\cite{Boyle:2019kee}. Meanwhile PN
  parameters are attributed to point particles, via asymptotic
  matching in ``body zones'', as explained
  in~\cite{poisson2014gravity}.  The difference between the PN and NR
  definitions should be small as we see in the optimization results,
  but perfect agreement between them is anticipated only at infinite
  separation, even for infinite-order PN and infinite-resolution NR.
  The leading order contribution to the difference is expected to be the binding energy between the two black holes, which appears at 1 PN order.
  Determining the exact relationship between PN and NR definitions requires solving for the apparent horizon in the PN metric and evaluating the parameters at the PN apparent horizon, which is beyond the scope of this paper. 
  We are working on this issue as a separate project.

  The difference in parameter definitions is not an error or uncertainty, but a conventional discrepancy, thus it has no physical significance. 
  For astrophysical parameter inference, it is recommended to quote PN parameters rather than NR, since these correspond to masses and spins infinitely far
in the past.

\item %
  Even if we used the same definitions of mass and spin quantities in
  PN and NR, the definitions are gauge dependent.
  Apparent horizon
  integrals used in NR are gauge dependent since apparent horizons are
  themselves slicing-dependent~\cite{Baumgarte_Shaprio:NumRel}.
  Meanwhile specifying a spin in PN requires a choice of spin
  supplementary condition (SSC), which is a type of gauge
  choice~\cite{Steinhoff:2015ksa}.  A choice of SSC also affects the
  definition of mass in PN.
  This gauge-dependence also applies to the total binary mass, which
  is distinct from the gauge-invariant ADM energy; these differ by the
  binding energy and the energy carried away by radiation.

\item Both PN and NR waveforms have systematic errors.  The PN errors
  are mainly due to the truncation at a finite PN order, and the NR
  errors are mainly due to numerical truncation.  The error is
  supposed to affect both the waveform and the parameters.  As we will
  show in Sec.~\ref{sec:Comparison}, the difference between the PN and
  NR parameters are at the same order as the residual between the
  strain modes.  Therefore, there is no reason to force the difference
  between the PN and NR parameters to be zero, seeing as this would
  introduce more error in the strain and thus make the hybrid waveform
  less accurate.  Given the finite length and finite resolution of the
  waveforms, the optimized parameters can compensate for the errors in
  the model and the unmodeled effects, to provide the best agreement
  and smoothest transition between the NR and PN waveforms.
\end{enumerate}

Therefore, we must identify the proper PN parameters that correspond to the NR systems.

Here we are concerned with optimizing only 12 parameters: 8 PN parameters ($q,M,\bm{\chi}_1$ and $\bm{\chi}_2$) and 4 coordinate parameters that determine the $\mathrm{SO}(3)$ rotation and time translation.  For technical reasons discussed in Sec.~\ref{sec:Procedure}, we use a different method to determine the additional supertranslation and boost parameters. We will present the complete procedure to determine all the parameters in Sec.~\ref{sec:Procedure}.

We determine these 12 parameters by minimizing the normalized $L^2$
difference over a chosen matching window between the fixed NR waveform
$h^{\mathrm{NR}}(t)$ and the free PN waveform,
$h^{\mathrm{PN}}(t;\vec{\lambda})$, where $\vec{\lambda}$ are the chosen parameters.  We define a cost function $\mathcal{E}$ from the normalized $L^2$ difference between the waveforms,
\begin{equation}
\mathcal{E}[h^{\mathrm{NR}}, h^{\mathrm{PN}}]=\frac{1}{2} \frac{\sum_{\ell, m} \int_{t_1}^{t_2}\left|h^{\mathrm{NR}}_{\ell m}(t) - h^{\mathrm{PN}}_{\ell m}(t;\vec{\lambda})\right|^2 d t}{\sum_{\ell, m} \int_{t_1}^{t_2}\left|h^{\mathrm{NR}}_{\ell m}(t)\right|^2 d t}
\,.
\label{eq:epsilon}
\end{equation}
Here $t_1$ and $t_2$ are the start and end time of the matching window.
$\mathcal{E}$ reduces to sphere-weighted average of the commonly used time-domain mismatch with a flat noise power spectral density (see Appendix C of~\cite{Blackman:2017dfb}).


Unless otherwise specified, we typically use a matching window ending at $-8000\,M$ for precessing systems, and $-5000\,M$ for spin-aligned systems, for the purposes of illustration. 
We set $t=0$ at the merger, where the merger is defined as the peak of the $L^2$ norm of the strain modes.

\begin{figure}[tb]
\centering
\includegraphics[width=\onecol]{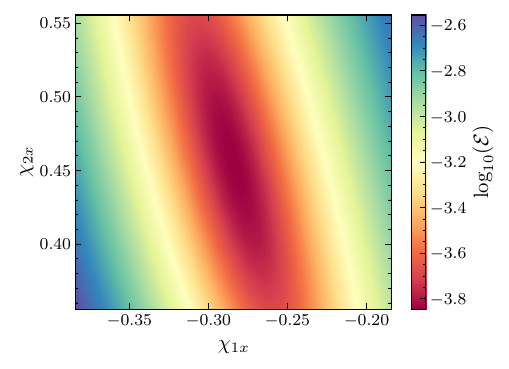}
\includegraphics[width=\onecol]{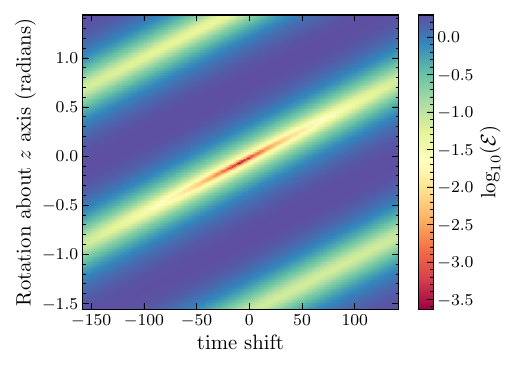}
\caption{%
  The profile of the cost function $\mathcal{E}$ with respect to
  parameters for system SXS:BBH:2617. The matching window is 20
  orbits long, and ends $8000\,M$ before
  merger. In each
  plot, the 10 other parameters are fixed to their
  best-fit values. The slanting long and narrow valleys show the
  degeneracy between these parameters, which can drastically slow down
  the optimization.}
\label{fig:CostProfile}
\end{figure}

We use the nonlinear least squares method to perform the
minimization.\footnote{%
  For the optimization, we use
  \codename{scipy.optimize.least\_squares} with the Trust Region
  Reflective algorithm~\cite{2020SciPy-NMeth, STIR}. The tolerances
  are set to $\text{gtol}=10^{-8}$ and $\text{xtol,
    ftol}=3\times10^{-15}$. We use the NR value of the parameters as
  the initial guess. The gradient is computed numerically. We impose bounds on the optimization parameters, which we will discuss later in the text.
}
However, it turns out that the cost function Eq.~\eqref{eq:epsilon} has long and narrow valleys that cause the optimizer to run slowly. 
Fig.~\ref{fig:CostProfile} shows two examples of the valleys in the
cost function. 
The top plot shows an example of the dependence of the cost function on the $x$ components of $\bm{\chi}_1$ and $\bm{\chi}_2$. 
We can see that the constraint on $\bm{\chi}_1$ is better than $\bm{\chi}_2$, although they still show some degeneracy between each other.
Additionally, the figure on the bottom illustrates the cross section involving the time shift and the $z$ component of the logarithm of the quaternion rotor that rotates the inertial frame.
This figure shows the degeneracy between time and phase.

\begin{figure}[tb]
\centering
\includegraphics[width=\onecol]{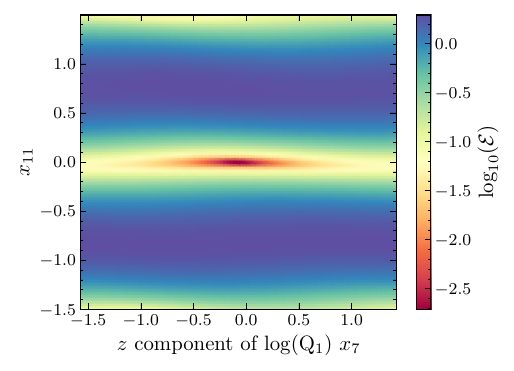}
\includegraphics[width=\onecol]{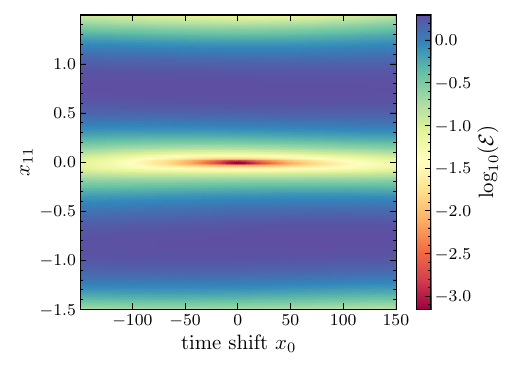}
\caption{The profile of the cost function $\mathcal{E}$ with respect to the new parameters for system SXS:BBH:2617. The matching window is 20 orbits long, and ends at $8000\,M$ before merger. In each plot, the 10 parameters that are not varied are fixed at the best-fit values. These new parameters are no longer degenerate. If the cost function has a narrow diagonal valley, as in Fig.~\ref{fig:CostProfile}, the optimizer tends to zigzag within the valley,
  leading to a significant slowdown in the optimization process. However, if the valley aligns with an axis, the optimizer can easily re-scale the axis, leading to more efficient optimization.}
\label{fig:CostProfileNew}
\end{figure}

In order to mitigate these degeneracies and enhance the optimization process, we reparametrize the 12 parameters to diagonalize the valleys.
First, we introduce new parameters $x_1$ to $x_4$ to describe the fractional change in PN mass ratio $q$, total mass $M$, and spin amplitudes compared to their NR counterparts; see Eq.~(\ref{eq:para1}).
Then, note that the NR value of the spin vectors could be measured in a different Euclidean frame from the PN frame, which is also suggested by the degeneracy between $\bm{\chi}_{1x}$ and $\bm{\chi}_{2x}$ shown in the upper plot of Fig.~\ref{fig:CostProfile}. We therefore introduce the quaternion $\bm{\mathrm{Q}}_1$ to describe the
rotational transformation of the two spin vectors as a unified entity.
This quaternion will have three degrees of freedom, which we parameterize as $(x_5,x_6,x_7)$.
One degree of freedom corresponds to a rotation \emph{about} $\bm{\chi}_{1}^\text{NR}$, and thus has no effect on $\bm{\chi}_{1}^\text{NR}$ itself. 
But if the two spins are not parallel, that degree of freedom can affect $\bm{\chi}_{2}^\text{NR}$.
This leaves just a single degree of freedom representing the rotation of $\bm{\chi}_{2}^\text{NR}$, in a direction orthogonal to both $\bm{\chi}_{1}^\text{NR}$ and $\bm{\chi}_{2}^\text{NR}$. 
This final rotation is encoded as another quaternion $\bm{\mathrm{Q}}_2$ that depends on only one additional parameter $x_8$.
Thus we have
\begin{equation}
\begin{aligned}
q &= x_1 q^\text{NR},\\
M &= x_2 M^\text{NR},\\
\bm{\chi}_1 &= x_3 \bm{\mathrm{Q}}_1 \bm{\chi}_{1}^\text{NR} \bm{\mathrm{\bar{Q}}}_1,\\
\bm{\chi}_2 &= x_4 \bm{\mathrm{Q}}_2\bm{\mathrm{Q}}_1 \bm{\chi}_{2}^\text{NR} \bm{\mathrm{\bar{Q}}}_1\bm{\mathrm{\bar{Q}}}_2,\\
\bm{\mathrm{Q}}_1 &= \exp\left[(x_5,x_6,x_7)\right],\\
\bm{\mathrm{Q}}_2 &= \exp\left[\frac{x_8}{2}\widehat{\left(\bm{\chi}_{1}^\text{NR}\times\bm{\chi}_{2}^\text{NR}\right)}\right],\\
\end{aligned}
\label{eq:para1}
\end{equation}
where the bar represents the conjugate of a quaternion, and the hat over the cross product indicates that the result is normalized.  ``NR'' superscripts denote the NR values of these quantities obtained at the reference time, which we typically set as the end time of the matching window.
Here we use the standard quaternion notation, where the exponential of
a 3-vector gives a unit quaternion~\cite{Boyle:2013nka}.

We also need a quaternion $\bm{\mathrm{R}}$, which describes an overall $\mathrm{SO}(3)$ rotation of the system,
\begin{equation}
\begin{aligned}
\bm{\mathrm{R}} &= \exp\left[(x_9,x_{10},x_{11})+\frac{x_{0}}{2}\bm{\Omega}_\text{tot}\right].
\end{aligned}
\label{eq:para2}
\end{equation}
The $\frac{x_{0}}{2}\bm{\Omega}_\text{tot}$ term in the quaternion
rotor $\bm{\mathrm{R}}$ is introduced to eliminate the degeneracy
between time and phase shown in the bottom plot in
Fig.~\ref{fig:CostProfile}. Here $x_0$ is the parameter that describes
the time shift between the NR and PN systems, and
$\bm{\Omega}_\text{tot}$ is the total angular velocity measured at the
reference time, which captures both orbital and precession contributions.
$\bm{\Omega}_\text{tot}$ is measured from the NR strain by defining it
as the angular velocity of a rotating frame in which the time
dependence of the waveform mode is minimized~\cite{Boyle:2013nka}.

When performing the nonlinear least squares optimization, we impose bounds on the parameters defined in Eqs.~\eqref{eq:para1} and~\eqref{eq:para2}.
For $x_{1-4}$, the bounds are windows of size $\pm0.05, \pm0.02, \pm0.01$ and $\pm0.01$ around their respective initial guesses.
For $x_{5-11}$, the range is $\pm\pi/4$, and $\pm0.5\pi/\Omega_{\text{orb}}$ for the time shift $x_0$.

Fig.~\ref{fig:CostProfileNew} shows how the new reparameterization can diagonalize the valleys of the cost function and mitigate the degeneracies. 
This parametrization has been tested on many systems. 
\begin{figure}[tb]
\centering
\includegraphics[width=\onecol]{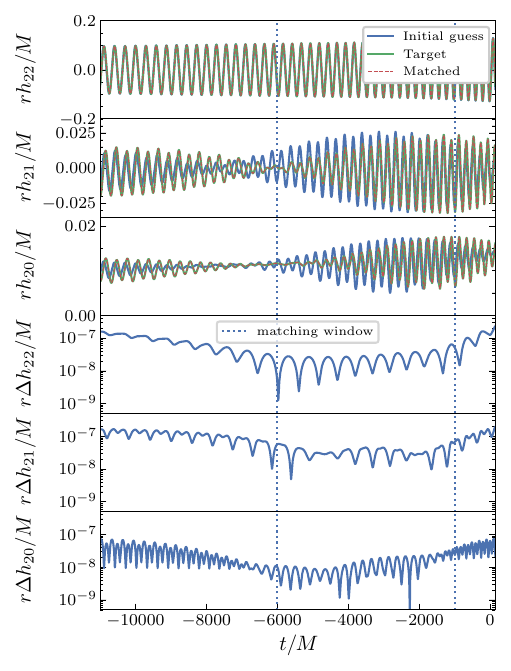}
\caption{Strain and residual of two PN waveforms, for a test in which we match the two waveforms by varying the parameters of one of them. The target PN waveform has parameters $q=1.5,\,M=1.0,\,\bm{\chi}_1=(0.5,0.2,0.3),\,\bm{\chi}_2=(0.1,0.3,0.4)$, and is rotated away by $\bm{\mathrm{R}}=\exp(0.05,0.02,0.03)$ from the original  frame in which the PN waveform is generated. The initial guess is $q=1.65,\,M=1.05,\,\bm{\chi}_1=(0.42,0.24,0.47),\,\bm{\chi}_2=(0.01,0.20,0.41)$. The parameters obtained by the optimization are $q=1.500004,\,M=1.000005,\,\bm{\chi}_1=(0.10001,0.200002,0.30003),\,\bm{\chi}_2=(0.399994,0.09998,0.20006)$. The matching window is the region between two vertical dotted lines, which is $5000\,M$ long.}
\label{fig:PNPNTest}
\end{figure}

\begin{figure}[tb]
\includegraphics[width=\onecol]{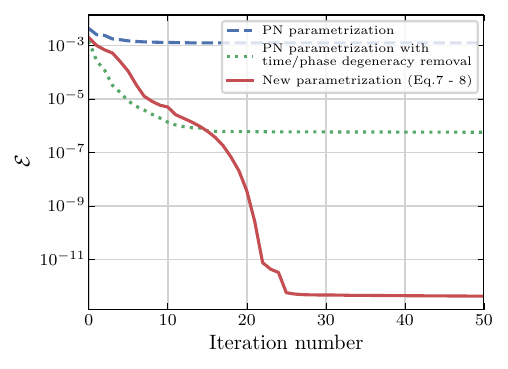}
\caption{The convergence of different parametrization methods for a PN-PN matching test, in which we match the two waveforms by varying the parameters of one of them. The target PN waveform has parameters $q=1.5,\,M=1.0,\,\bm{\chi}_1=(0.5,0.2,0.3),\,\bm{\chi}_2=(0.1,0.3,0.4)$, and is rotated away by $\bm{\mathrm{R}}=\exp(0.0,0.0,0.05)$ from the original frame in which the PN waveform is generated. The initial guess is $q=1.65,\,M=1.05,\,\bm{\chi}_1=(0.42,0.24,0.47),\,\bm{\chi}_2=(0.01,0.20,0.41)$. The matching window is $5000\,M$ long. The result using the new parametrization in Eqs.~\eqref{eq:para1} and~\eqref{eq:para2} is shown as the red curve, which converges much faster and better than the original PN parametrization, which is shown as the blue dashed curve.
  The green dotted curve also uses the original PN parametrization but incorporates the $\frac{x_{0}}{2}\bm{\Omega}_\text{tot}$ term in the frame rotor $\bm{\mathrm{R}}$ defined in Eq.~\eqref{eq:para2}. In other words, it uses ($x_0$, $q$, $M$, $\bm{\chi}_1$ and $\bm{\chi}_2$) and ($x_9$ through $x_{11}$) as parameters. The green dotted curve shows better convergence than the blue dashed curve.}
\label{fig:OptIter}
\end{figure}

We test the optimization routine by using a PN waveform with known parameters as a proxy ``NR'' waveform, and adjusting the parameters of another PN waveform until the two waveforms match.
In most cases, both the original PN parametrization and our new parametrization presented in Eq.~\eqref{eq:para1} and \eqref{eq:para2} can recover the PN parameters to high accuracy and reduce the waveform differences to $\mathcal{E} \approx 10^{-15}$, as shown in Fig.~\ref{fig:PNPNTest}.
But for poor initial guesses, our new parametrization generally behaves better than the original PN parametrization, as shown in Fig.~\ref{fig:OptIter}.

\section{Hybridization procedure}
\label{sec:Procedure}

\begin{figure}[tb]
\includegraphics[width=\onecol]{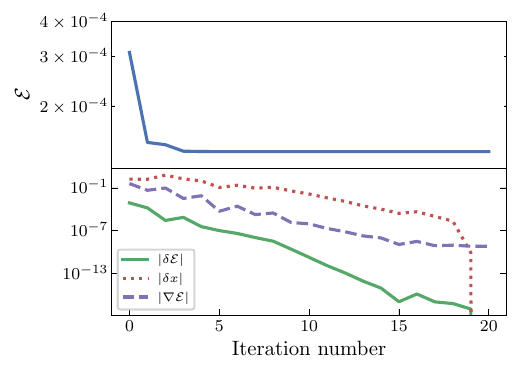}
\caption{%
  The convergence of the 12-D nonlinear least-squares optimization for
  system SXS:BBH:2617 in the second iteration of step \ref{item:iter}
  of the procedure in Section~\ref{sec:Procedure}, using the
  parameters from the first iteration as the initial guess. The
  horizontal axis is the iteration number for the 12-D
  optimization. The end of the matching window is $5000\,M$
  before merger, and is 20 orbits long. The blue curve shows the value
  of the cost function, and the green solid and red dotted curves show the
  reduction of the function value and the norm of change in parameters
  after each iteration. The norm of the gradient of the cost function
  is shown in dashed purple.
}
\label{fig:OptConvergenceNR}
\end{figure}

\begin{figure}[tb]
\includegraphics[width=\onecol]{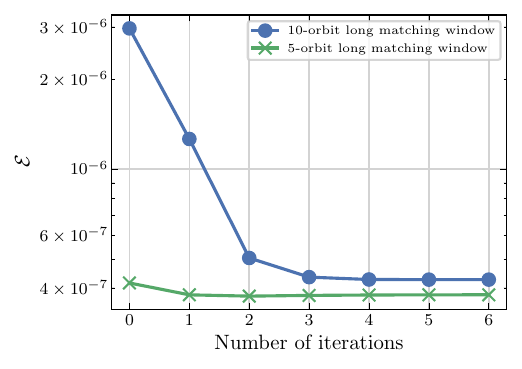}
\caption{%
  The convergence of iterations of step \ref{item:iter} of the
  procedure in Section~\ref{sec:Procedure} for system SXS:BBH:1414. The
  0th iteration is step \ref{item:PN-opt0} of the procedure in
  section~\ref{sec:Procedure}, for which the NR waveform is mapped
  into the superrest frame, and for this iteration the error does not
  include the $m=0$ modes since the value of the strain in these modes
  in the superrest frame does not agree with PN. For the other
  iterations, the NR waveform is mapped to the PN BMS frame, and the
  errors include the $\ell\le 8$ modes. The blue curve
  uses a 10-orbit long matching window that ends $8000\,M$ before
  merger, and the green curve uses a 5-orbit long matching window
  that ends $2000\,M$ before merger. Both curves converge within 4
  iterations. The cost function for the shorter matching window is
  smaller and converges less rapidly than the case with a longer
  matching window, likely because the short waveforms are being
  overfit.  This overfitting is explored in
  Sec.~\ref{sec:ShortWaveform}.}
\label{fig:IterConverge}
\end{figure}

In the previous section we described how we determine the PN parameters, the time offset, and the overall rotation of a PN waveform that we match to an NR waveform. 
However, we also need to account for differences in supertranslations and boosts between the NR and PN waveforms. 
While it is possible to add extra supertranslation parameters into the cost function defined in the previous section, we find it more efficient to account for all parameters (8 PN physical parameters, time shift, boost, rotation, and supertranslations) using the following iterative procedure: 

\begin{enumerate}
\item \label{item:superrest} Map the NR waveform to its superrest frame, which is defined as the frame in which the Moreschi supermomentum is minimized \cite{Mitman:2022kwt}. The superrest frame is determined during inspiral at the middle of the matching window.

The original CCE waveforms are typically in an arbitrary BMS frame with large center-of-mass charge and supermomentum, resulting in significant oscillations in the strain mode amplitudes and orbital frequency.
These irregularities introduce bias when determining the PN parameters. 
Mapping the NR waveform to its superrest frame reduces the effects caused by a large center-of-mass charge and supermomentum, mitigating biases and enabling a more  accurate optimization of the PN parameters.

\item Set the $m=0$ modes of the NR and PN waveforms to zero in their corotating frame. 

The $m=0$ modes have the most memory contribution, and are thus more influenced by the supertranslations involved in the BMS frame transformation.
Since the superrest frame of the NR system differs from the PN BMS frame, the strain modes in these two frames are expected to differ as well.
This can also introduce bias when optimizing PN parameters.
Therefore, we set $m=0$ modes to zero in their corotating frame to exclude the principal memory contributions to the strain.
Note that the $m=0$ modes will be restored in step \ref{item:BMS-transf} and the steps thereafter.

\item \label{item:PN-opt0} Perform the 12D optimization described in the previous section to determine the 8 PN parameters, rotation, and time shift.

\item \label{item:BMS-transf} Now with the $m=0$ modes of the original NR waveform included, map the original NR waveform to the BMS frame of the resulting PN system by applying the supertranslations that map the NR Moreschi supermomentum to the PN Moreschi supermomentum, and also applying the boost that minimizes the center-of-mass charge. Note that the PN Moreschi supermomentum depends on the PN parameters, so the NR waveform obtained by this step depends on the current values of the PN parameters.

\item \label{item:PN-opt} Again perform the 12D optimization, but using the NR waveform from step \ref{item:BMS-transf}, including the $m=0$ modes in the PN waveform.
  
\item \label{item:iter} Iterate steps
  \ref{item:BMS-transf}
  and
  \ref{item:PN-opt}
  until the relative change in $\mathcal{E}$ is less than $10^{-2}$.
  
\item Use a smooth stitching function to hybridize the PN and NR waveforms over the matching window. 
\end{enumerate}

Our stitching function is the smooth transition function
\begin{multline}
\tau(t)= \\
\begin{cases}
  0 &  t<t_1, \\
  \left\{1+\exp\left[\left(t_2-t_1\right)\left(\frac{1}{t-t_1}+\frac{1}{t-t_2}\right)\right]\right\}^{-1} & t_1 \leq t \leq t_2, \\
  1 & t_{2} <t.
\end{cases}
\end{multline}
The hybridized waveform is then
\begin{equation}
h_{\ell m}^{\mathrm{Hyb}}(t) = [1-\tau(t)] h_{\ell m}^{\mathrm{PN}}(t)+\tau(t) h_{\mathrm{BMS}\,\ell m}^{\mathrm{NR}}(t),
\end{equation}
where $h^{\mathrm{PN}}$ is the PN waveform obtained with the given $x_i$ parameters and $h_{\mathrm{BMS}}^{\mathrm{NR}}$ is the NR waveform in the adjusted BMS frame, as obtained in step \ref{item:BMS-transf} of the iterative procedure.

The convergence of the 12-D optimization is tested for the NR-PN hybridization case with comparable BH masses, as shown in Fig.~\ref{fig:OptConvergenceNR}. 
The optimization typically converges within tens of iterations.

The process of iteratively optimizing PN parameters and mapping the NR waveform to the resulting PN BMS frame demonstrates rapid convergence for cases with comparable BH masses, as illustrated in Fig.~\ref{fig:IterConverge}.
In the case of shorter matching windows, the decrease in the cost function during each iteration is less pronounced compared to longer matching windows. 
This can be attributed to the limited information contained within shorter windows, which may result in overfitting.
We will further address this problem in Sec.~\ref{sec:ShortWaveform}.
The iteration process typically converges within 2--4 steps.

For cases with a larger mass ratio ($q\sim 8$), larger eccentricity ($e>0.05$), or larger $\chi_p$ ($>0.5$), both the 12-D optimization and the overall iterative procedure converge more slowly. 
For these cases, the 12-D optimization typically takes more than a hundred iterations, and the overall iterative procedure typically takes more than 10 iterations. 
This slowdown occurs because for systems with large mass ratios, the waveforms become less sensitive to the smaller black hole's properties.
And for systems with larger $\chi_p$ or eccentricity, the PN and NR waveforms that we use become intrinsically different, as we will show in Sec.~\ref{sec:error}.
These features make the optimization more challenging.

\section{Results}
\label{sec:results}

\begin{figure}[tb]
\centering
\includegraphics[width=\onecol]{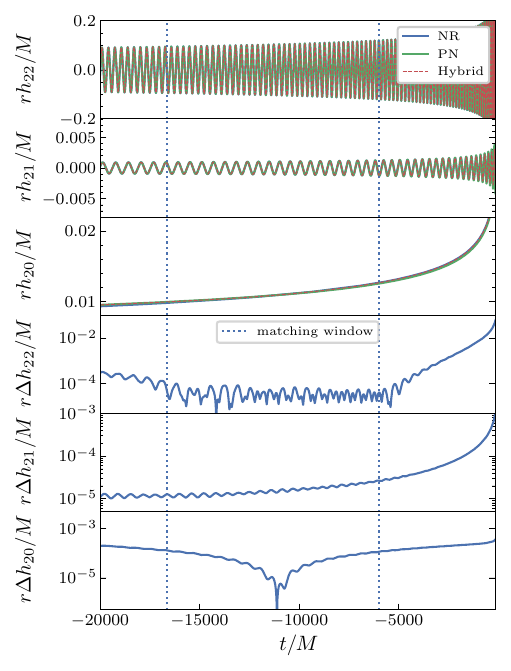}
\caption{Strain and residual of the hybridized spin-aligned system SXS:BBH:1412. The matching window is the region between the two vertical dotted lines, which is 20 orbits long and ends at $-6000\,M$. The blue, green and red curves in the upper panel are the strain modes of the NR, PN and hybrid waveforms. The blue curves in the lower panel are the residuals between the NR and PN strain modes.}
\label{fig:ResultSpinAlign}
\end{figure}

\begin{figure}[tb]
\centering
\includegraphics[width=\onecol]{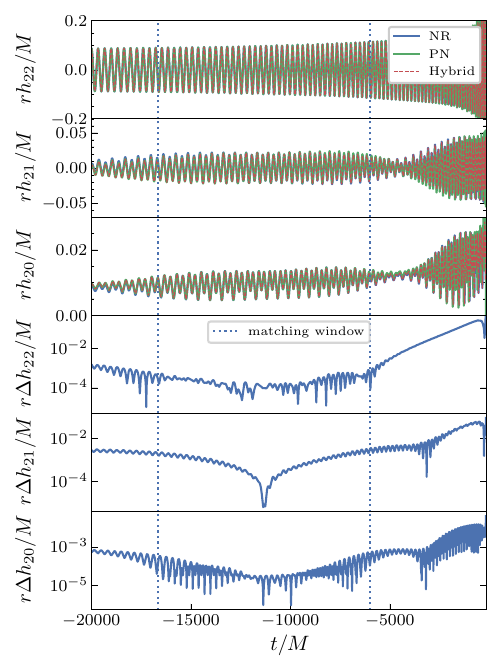}
\caption{Strain and residual of the hybridized precessing system SXS:BBH:2619. The matching window is the region between the two vertical dotted lines, which is 20 orbits long and ends at $-6000\,M$. The blue, green and red curves in the upper panel are the strain modes of the NR, PN and hybrid waveforms. The blue curves in the lower panel are the residuals between the NR and PN strain modes.}
\label{fig:ResultPrecessing}
\end{figure}

\subsection{Comparison between PN and NR waveforms}
\label{sec:Comparison}
We present typical hybridization results for spin-aligned systems and precessing systems in Figs.~\ref{fig:ResultSpinAlign} and~\ref{fig:ResultPrecessing}.
For the early inspiral stage, the PN and NR waveforms agree quite well, up to the systematic errors (eg., non-zero eccentricity of NR systems, absence of PN terms, etc.) that we will discuss in Sec.~\ref{sec:error}.

Apart from systematic errors in the early inspiral stage, we notice the PN failure thousands of $M$ before merger as the differences between the NR and PN strain modes become much larger.
We will be interested in the frequency at which the PN truncation error becomes larger than other systematic errors that affect our hybridization procedure.
Later we will choose the matching window to end before this frequency, so that the PN truncation error has a negligible effect on the matching procedure compared to the other errors.

The PN expansion parameter is $v=\left(M\Omega_\text{orb}\right)^{1/3}$, and each power of $v$ counts for $\tfrac{1}{2}$PN order.
PN waveforms that are complete up to $N$ PN order have truncation errors at relative order $\mathcal{O}\left(\Omega_\text{orb}^{(2N+1)/3}\right)$.
Therefore, the frequency at which PN truncation error becomes dominant satisfies
\begin{equation}
  \left(\text{Coefficient}\cdot\Omega_\text{orb}^{(2N+1)/3}\right)^2 \approx
  \mathcal{E}_{\mathrm{sys}},
\label{eq:failure}
\end{equation}
where $\mathcal{E}_{\mathrm{sys}}$ is the norm of differences induced by other systematic errors.
The systematic errors include errors that can cause the differences between NR and PN beyond PN truncation. These may arise from numerical errors, eccentricities in NR binaries that are meant to be quasicircular, inaccuracies in PN parameters, imprecise BMS frame alignment, incomplete consideration of certain physical aspects in PN or NR models like high order spin interactions in PN, or inaccurate boundary conditions in NR, and more. We will explore further the systematic errors that limit our hybridization accuracy in Sec.~\ref{sec:error}.

\subsection{Improvement from PN parameters and BMS frame fixing}
\label{sec:Improvement}
\begin{figure*}[tb]
\includegraphics[width=\twocol]{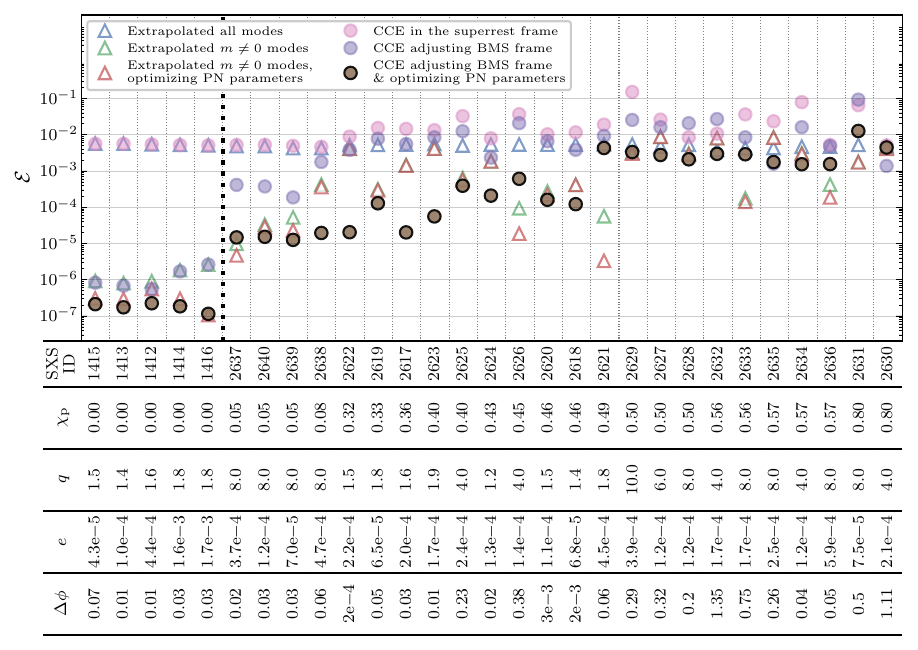}
\caption{Hybridization errors in the matching window, Eq.~(\ref{eq:epsilon}),
for different waveforms and for different methods of choosing PN parameters and
matching the BMS frame. See text for details. All the errors are measured after
adjusting the 3-D rotation and time shift. The BMS frame for extrapolated
waveforms cannot be modified using the method described in this paper because
those waveforms do not have an accurate value of $\Psi_2$ to calculate the
Moreschi supermomentum. We only adjust the BMS frame for CCE waveforms. Systems
on the left of the vertical dotted line are spin-aligned systems; systems on the
right are precessing systems, and systems are ordered left to right by $\chi_p$.
The mass ratios for SXS:BBH:2617--2635 systems are $1<q<10$ and for
SXS:BBH:2636--2640 systems are $q\approx8$. The matching window is 10 orbits.
For systems other than the SXS:BBH:2636--2640 systems, the matching window ends
$8000\,M$ before merger. For the SXS:BBH:2636--2640 systems the matching window
ends $5000\,M$ before merger because these systems are not long enough. The
brown markers use the full hybridization procedure described in
Section~\ref{sec:Procedure}. $e$ is the eccentricity measured from the trajectory at the NR reference time, and $\Delta \phi$ is the dephasing between the two highest resolution NR runs of the last 20 orbits before merger. It is important to recognize that this dephasing does not directly reflect the inaccuracy of the NR waveforms, as the NR parameters within the matching window for different Levs can deviate by $10^{-2}$, given their evolution over nearly a hundred orbits prior to the matching window.}
\label{fig:Improvement}
\end{figure*}

\begin{figure*}[tb]
\includegraphics[width=\twocol]{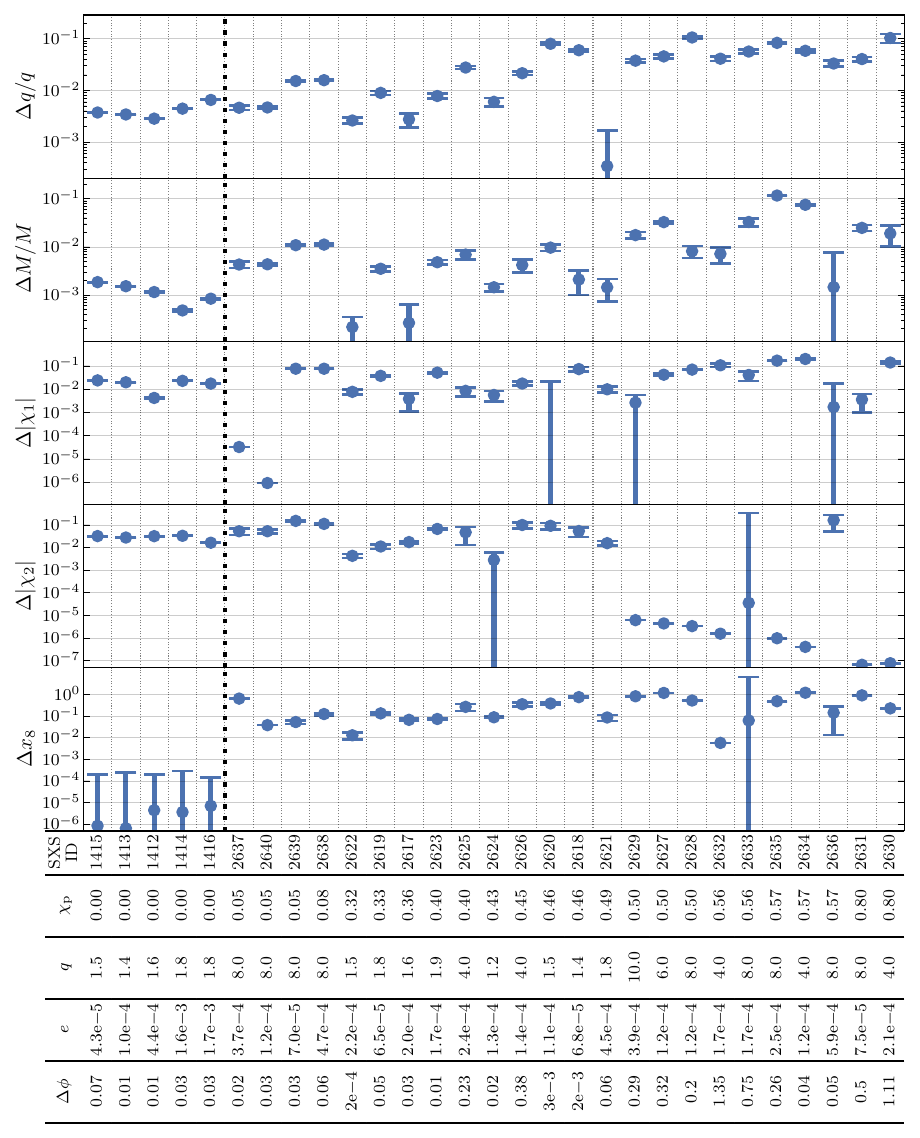}
\caption{The differences in parameters between NR and PN for selected waveforms
hybridized using the procedure of  Section~\ref{sec:Procedure}. Plotted $\Delta
X$ is $|X_\mathrm{NR}-X_\mathrm{PN}|$ for selected parameters $X$. Systems on
the left of the vertical dotted line are spin-aligned systems; systems on the
right are precessing systems, and systems are ordered left to right by $\chi_p$.
The mass ratio for SXS:BBH:2617--2635 systems are $1<q<10$, for
SXS:BBH:2636--2640 systems are $q\approx8$. The error bars show the
uncertainties of the optimization, and the markers show the best-fit value of
the parameters.}
\label{fig:Parameter}
\end{figure*}

\begin{figure}[tb]
\includegraphics[width=\onecol]{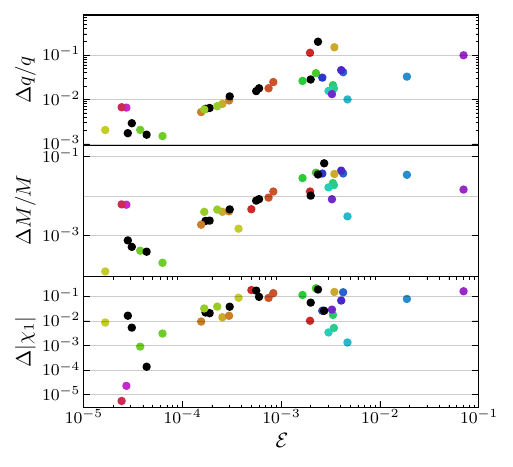}
\caption{The differences in parameters between NR and PN with respect to $\mathcal{E}$ in a 10-orbit long matching window. Each color represents a different system. The points with the same color represent the same system with different matching windows. All the systems in this figure are precessing systems.}
\label{fig:ParameterE}
\end{figure}

In Figure~\ref{fig:Improvement}, we present the hybridization errors
of 29 systems. The hybridization error plotted is the cost function in Eq.~\eqref{eq:epsilon} computed within a 10-orbit matching window.
However, only the brown markers in the figure represent errors computed using the full hybridization procedure described in Section~\ref{sec:Procedure}. 
The other markers represent results obtained using only parts of the procedure, omitting some of the waveform modes, or using NR waveforms that are not obtained using CCE; these markers are included to illustrate the need for all parts of the procedure described in Section~\ref{sec:Procedure}.
For all the markers, the 4 parameters representing the 3D rotation and time shift between the PN and NR waveforms are optimized to obtain a minimal hybridization error.
The triangular markers use extrapolated NR waveforms obtained using the methods described in \cite{Boyle:2009vi},
which lack the correct memory effect.
The circular markers use CCE waveforms that include memory contributions and contain enough information to allow suitable BMS transformations.
The blue triangular markers denote errors obtained from the extrapolated waveforms without adding supertranslations, and with the 8 PN parameters fixed to their corresponding NR values. 
The extrapolated waveforms get a different CoM-correction that is described in \cite{Woodford:2019tlo}.
Because the extrapolated waveforms lack the correct memory terms, we include the green triangular markers, which are obtained by setting the $m=0$ modes to zero in the NR corotating frame before hybridization, and then plotting the hybridization error of only the $m\neq0$ modes. 
For the green triangular markers, the PN parameters are fixed to their corresponding NR values.
The errors represented by the pink, purple, and brown circular markers include all modes including $m=0$.
The pink circular markers utilize CCE waveforms, but the NR waveforms are in their inspiral superrest frames, and the PN parameters are fixed to their corresponding NR values at the reference time.
The purple circular markers map the NR waveforms to the corresponding PN BMS frame, but the PN parameters used in both the PN waveform and to define the PN BMS frame are the corresponding NR values of those parameters.
Comparing the results in the superrest frame against those in the PN BMS frame (pink and purple circular markers), we observe that by fixing the BMS frame we can improve the error between PN and NR waveforms by 1 to 3 orders of magnitude. 
This indicates that CCE NR waveforms are usually in a BMS frame that differs from the PN BMS frame, and aligning the BMS frame is crucial when comparing or hybridizing two GW waveforms.

By optimizing PN parameters in addition to optimizing time shift and rotation, as shown in the red and brown, we further reduce $\mathcal{E}$ by 2 to 3 orders of magnitude for spin-aligned systems and up to 3 orders of magnitude for precessing systems. 
This suggests that the NR parameters at the reference time are different from the parameters that yield the best-matching PN waveform.
Figure~\ref{fig:Parameter} shows the difference in parameters for selected systems, using the full hybridization procedure described in Section~\ref{sec:Procedure}, which corresponds to the brown markers in Fig.~\ref{fig:Improvement}.
We see that the fractional difference in mass ratio $q$ between PN and NR is around
$10^{-2}$, the fractional difference in total mass $M$ is around $10^{-3}\sim10^{-2}$, and
the absolute difference in spin magnitudes is around $10^{-2}\sim10^{-1}$.
Fig.~\ref{fig:ParameterE} shows the correlation between the PN/NR parameter difference and $\mathcal{E}$ in the matching window.
Note that $\mathcal{E}$ is the $L^2$ error, so the residual in the strain is about $\mathcal{E}^{1/2}$, which is around the same order as the difference in the parameters, as expected.

Given that the best-fit PN parameters can be significantly different from the NR parameters in some cases, it is possible that the optimization over intrinsic PN
parameters leads to overfitting. 
Missing PN terms or NR truncation errors can potentially be absorbed by this optimization, leading to an incorrect PN waveform getting attached to the NR waveform.
While Figure~\ref{fig:Improvement} does not capture this effect because it only shows the errors in the matching window, one can check this by comparing the errors between PN and NR in an earlier test window that was not included in the optimization. 
This will be done in Sec.~\ref{sec:WindowChoice}.

\section{Sources of error}
\label{sec:error}

In this section, we study the origin of the hybridization error. 
We have two goals in mind: first, we would like to better understand how to improve the hybridization procedure using current NR waveforms and currently known PN terms. 
Second, we would like to know whether the hybridization error is at present limited by our current NR waveforms or our knowledge of PN, so that we can understand what is needed in terms of NR or PN improvements to reduce hybridization errors in the future.

\subsection{Spin-aligned systems}

\label{sec:window}
\begin{figure}[tb]
\includegraphics[width=\onecol]{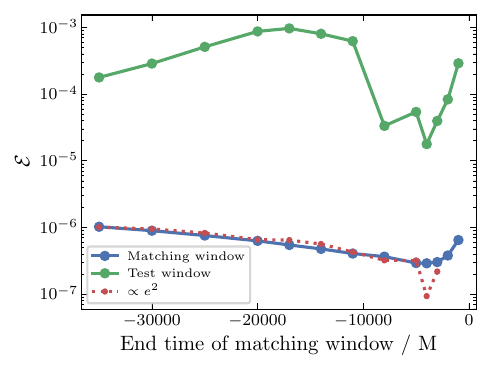}
\caption{Hybridization error for different locations of the matching
  window for system SXS:BBH:1412. We vary the end time of the
  matching window, while keeping the matching window 15 orbits
  long. The test window is 10 orbits long, and the center
  (in number of orbits)
  of the test window is 30 orbits
  earlier than the center of the matching window. The red dotted curve shows
  the square of the eccentricity measured using \codename{gw\_eccentricity} \cite{Shaikh:2023ypz} by the amplitude fitting method at the middle of the
  matching window multiplied by a factor of 15. The factor is obtained
  by varying the overall amplitude of the red curve until it overlays the blue curve.
\label{fig:Window}}
\end{figure}

For spin-aligned systems, we can see from Fig.~\ref{fig:Improvement} and Fig.~\ref{fig:Window} that the error inside the matching window is on the order of $10^{-7}$.

The blue curve in Fig.~\ref{fig:Window} shows $\mathcal{E}$ in a 15-orbit long matching window with a varying end time, and it has a local minimum when the angular velocity is around 0.012, or between 3000--8000 $M$ before merger. 
The increase in hybridization error after the minimum, where the end time of the matching window gets closer to merger,
corresponds to the expected late-time PN failure that we discussed in Sec.~\ref{sec:Comparison}.
We will show below that the increase in hybridization error before the minimum, where the matching window is moved earlier and thus farther from merger, is consistent with the effects of using an NR waveform with small but nonzero orbital eccentricity to hybridize with a PN waveform that assumes eccentricity is identically zero.
The test window error (green curve) in Fig.~\ref{fig:Window} will be discussed later in Sec.~\ref{sec:window}.

The contribution of eccentricity to the leading order in PN quasi-Keplerian expansion is given by \cite{Ebersold:2019kdc}
\begin{equation}
\label{eq:ecc}
h_{\text {Newt }}^{22} \propto 1+e\left(\frac{1}{4} \exp(-\mathrm{i} \xi)+\frac{5}{4} \exp(\mathrm{i} \xi)\right),
\end{equation}
where $e$ is the time-evolving eccentricity parameter.  The angle $\xi$ is closely related to the mean anomaly, going through roughly $2\pi$ with each orbit.
From Eq.~\eqref{eq:ecc} we can obtain the PN prediction of the $L^2$ normalized difference $\mathcal{E}_{\mathrm{sys}}$ between eccentric PN waveforms and non-eccentric PN waveforms in the limit of small eccentricity as
\begin{equation}
\mathcal{E}_\mathrm{sys}\propto e^2.
\label{eq:EpsilonE}
\end{equation}
\begin{figure}[tb]
\includegraphics[width=\onecol]{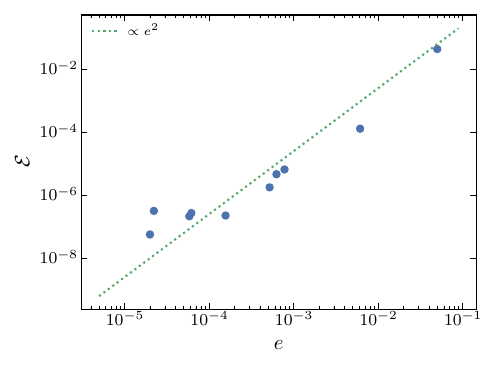}
\caption{The $L^2$ normalized error $\mathcal{E}$ versus eccentricity $e$ for 10 different systems: SXS:BBH:1412--1416, SXS:BBH:1153, 1164, 1165, 1168; and the additional system 3618.
We include SXS:BBH:3618 for reference as a higher-eccentricity point ($e=0.0496$).  It is spin-aligned with $\chi_1=\chi_2=0.6$ and $q=1.0$, and will be added to the public SXS catalog soon.
The end time of the matching window is $-5000\,M$, and the length of the matching window is 10 orbits. The green dotted curve is a quadratic fit to the plotted points.}
\label{fig:EpsilonE}
\end{figure}%
In Fig.~\ref{fig:EpsilonE} we plot $\mathcal{E}$ between PN and NR for 10 different systems. 
The error increases with the eccentricity, and is roughly consistent with the expected $\mathcal{E}\propto e^2$ behavior of Eq.~\eqref{eq:ecc}.

In addition to examining the overall magnitude of $\mathcal{E}$ caused by eccentricity, we also examine the time evolution of eccentricity and its effect on $\mathcal{E}$. 
The time evolution of eccentricity to leading order is given by \cite{Moore:2016qxz}
\begin{equation}
\label{eq:eccEv}
e_t(\Omega_t)\approx e_0\left(\frac{\Omega_0}{\Omega_t}\right)^{19/18},
\end{equation}
where $\Omega_t$ is the orbital frequency at time $t$, and $\Omega_0$ is the orbital frequency at a fiducial time when the eccentricity is $e_0$.
According to Eqs.~\eqref{eq:ecc} and \eqref{eq:EpsilonE}, the oscillation amplitude of the norm of the $(2,2)$ mode should be proportional to $e\propto \Omega^{-19/18}$.
We check the oscillation amplitude of the norms of the $(2,2)$, $(2,-2)$, $(2,1)$ and $(2,-1)$ modes for system SXS:BBH:1412. 
The oscillation amplitude is obtained by subtracting the individual mode's norm from its orbital average, which is obtained by applying a low-pass filter to the original mode.
We find that the oscillation amplitudes of the norms of the above modes are proportional to $\Omega^{-19/18}$ and are thus consistent with eccentricity damping.
The same argument applies to $\mathcal{E}$ within the matching window as we show in Fig.~\ref{fig:Window}.
Assuming that the change of $e$ and $\Omega$ in the window is small, then $\mathcal{E}\propto e^2 \propto \Omega^{-19/9}$.
The red dotted line is $e^2$ multiplied by a coefficient chosen to
approximately agree with the matching window error (blue curve), where the eccentricity $e$ is measured from the waveform amplitude fit of the $(2,2)$ mode using the package \codename{gw\_eccentricity} \cite{Shaikh:2023ypz}, and measurements are taken at the middle of the different matching windows.
The value of $\mathcal{E}$ in the matching window (blue curve) agrees very well with the trend of $e^2$ (green curve), showing the expected effect of eccentricity damping.

We also remark, based on Eq.~\eqref{eq:ecc}, that the eccentricity contribution oscillates at the same frequency as the orbital frequency, and is symmetric for positive and negative $m$. 
We therefore check the oscillation amplitudes of the norms of the $(2,2)$ and $(2,-2)$ modes for a system with a greater eccentricity, SXS:BBH:1168. 
The oscillation amplitudes and phase for these two modes are indeed the same, which is consistent with the above arguments for eccentricity.

\begin{figure}[tb]
\includegraphics[width=0.47\textwidth]{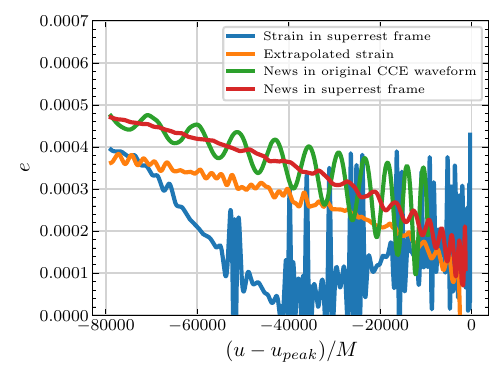}
\caption{%
  Measurement of eccentricity $e$ from both strain and news (the time
  derivative of strain), and in different BMS frames, for system
  SXS:BBH:1412.  When transforming to a superrest frame, the frame
  is determined from the middle of the inspiral.}
\label{fig:ESupertranslate}
\end{figure}
On the basis of these arguments, one might assume that the hybridization error would be reduced if we included eccentricity and mean anomaly in our PN parameter optimization. 
However, by conducting such a 14-dimensional optimization in the superrest frame of the NR system, we found that the error couldn't be reduced further since the error generated by such a small eccentricity is overpowered by the error caused by the incorrect BMS frame.
Fixing the BMS frame, however, is not a simple task when eccentricity is present, as both eccentricity and supertranslation can produce oscillations in the waveform mode amplitudes and the angular velocity. 
This degeneracy is depicted in Fig.~\ref{fig:ESupertranslate}, where we plot the eccentricity measurement in different BMS frames as a function of time.
We can see by comparing the eccentricity measurement using a CCE waveform in the superrest frame (blue) and an extrapolated waveform (orange) that the measurement of eccentricity from the strain has a strong dependence on the method of obtaining the asymptotic waveform.
The measurement of eccentricity from the Bondi news---the time derivative of the strain---is less dependent on the BMS frame, but there are still large oscillations with respect to the measurement time (green and red lines).
Because of the degeneracy between supertranslations and eccentricity, including eccentricity in hybridization is nontrivial and we leave it for future work.

\subsection{Precessing systems}
\label{sec:errorPrecessing}
\begin{figure}[tb]
\includegraphics[width=0.5\textwidth]{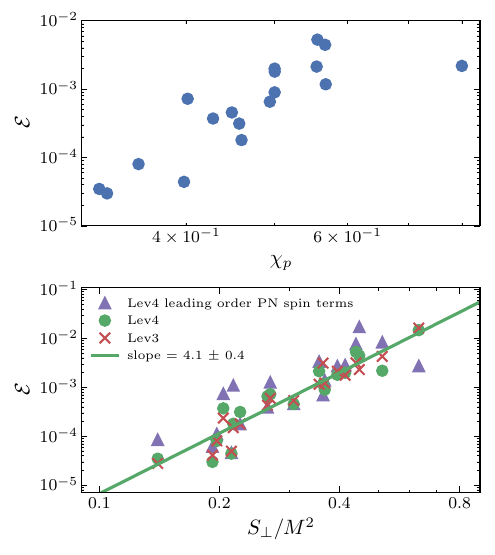}
\caption{Hybridization error, Eq.~(\ref{eq:epsilon}), inside the matching window for precessing systems SXS:BBH:2617--2635. The horizontal axis is the spin-precession parameter $\chi_p$ for the upper panel, and $S_{\perp}/M^2$ for the lower panel. Here $\bm{S}_i=\bm{\chi}_im_i^2$ is the spin angular momentum, and $S_\perp$ is the magnitude of the projection in the orbital plane. Results are shown for two different choices of the resolution parameter ``Lev'' used to compute the NR simulation. Lev4 uses a finer tolerance for the adaptive-mesh-refinement algorithm than Lev3, so Lev4 has smaller numerical errors. The additional purple markers use only the leading order spin-spin and spin-orbit terms in both strain and $\Psi_M$ for Lev4, whereas the other markers use all PN terms listed in Sec.~\ref{sec:PNwaveforms}, which have higher order terms in the strain, but the same terms in $\Psi_{\mathrm{M}}$. The matching window is 10 orbits long, and the end time is $8000\, M$ before merger.}
\label{fig:SourceErrorPrecessing}
\end{figure}

In Figure~\ref{fig:SourceErrorPrecessing}, we present the hybridization errors within the matching window for various precessing systems.
The horizontal axis for the upper panel is the spin-precession parameter $\chi_p$.
Notably, these blue dots reveal a correlation between the
hybridization error and the spin precession for each system.

To discern whether the error is caused by the numerical error of NR spin-precession or the absence of spin-precession terms in PN, it is important to recognize that the spin terms entering the PN expansion are the spin angular momentum $\bm{S}_i=\bm{\chi}_im_i^2$, rather than $\chi_p$.
To investigate this, in the lower panel we plot the hybridization error for the same systems against $S_{\perp}/M^2=(S_{1\perp}+S_{2\perp})/M^2$ using two different NR simulations that are identical except for the resolution parameter ``Lev'' that determines the error tolerance of the simulation.  
Larger ``Lev'' means a smaller error tolerance, which translates into a more accurate (but more computationally expensive) NR simulation.
The same Lev for all simulations used in this paper corresponds to the same tolerance for the adaptive-mesh-refinement algorithm. See the last paragraph of Sec. 2.1 in Ref.~\cite{Boyle:2019kee} for details.
By comparing the results for Lev4 and Lev3 (green dots and red crosses), it is evident that the NR numerical error is not the limiting error source. 
The slope of the Lev4 $L^2$ norm error (green dots) is $3.8$, implying that the slope of the absolute difference between PN and NR waveforms should be $1.9$, since the residual is around the square root of $\mathcal{E}$. Note that the contribution of the missing PN precession-dependent spin-spin terms to the waveform is expected to be proportional to $(S_{\perp}/M^2)^2$, so our result is consistent with those missing terms being the dominant source of error.

To verify consistency, we included results that only use the leading order spin-spin and spin-orbit terms in both the strain and $\Psi_M$ for Lev4 (purple triangular markers). It is important to recognize that this comparison is not fully consistent, as the Lev4 scenario represented by green dots also includes the spin-spin and spin-orbit PN terms at only the leading order in $\Psi_M$, but has up to the next-to-next leading order in the strain. Nevertheless, there are still differences when using different terms in the strain at higher values of $S_{\perp}/M^2$. The slope for markers including only the leading order spin terms is $3.6\pm0.6$, suggesting that higher order PN spin terms might influence the hybridization error.

\section{Choice of matching window}\label{sec:WindowChoice}

\begin{figure}[tb]
\includegraphics[width=\onecol]{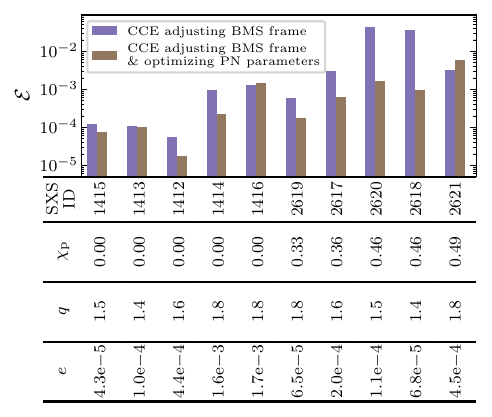}
\caption{Test window error for 10 long waveforms from
Fig.~\ref{fig:Improvement}. The center of the test window is 30 orbits earlier
than the center of the matching window. The matching window is the same with
Fig.~\ref{fig:Improvement}. For most systems, the optimization over PN
parameters can result in smaller error in both the matching window and the test
window.}
\label{fig:ImprovementTest}
\end{figure}

The matching window should be selected so that both PN and NR are valid there and agree with each other to an acceptable tolerance.
We can estimate the matching window's end time by Eq.~\eqref{eq:failure}, which estimates the frequency at which PN becomes inaccurate at late times: When the left-hand side of this equation exceeds the right-hand side, the PN truncation error becomes dominant in the hybridization error, and this error grows polynomially with the
orbital frequency as we get closer to merger.
If the matching window is chosen earlier so that the left-hand side is smaller than the right-hand side, the hybridization error is then dominated by the remaining systematic errors that we are not able to eliminate for now.

However, Eq.~\eqref{eq:failure} typically has undetermined coefficients, which depend on parameters of the system. 
In addition, Eq.~\eqref{eq:failure} does not tell us how to determine the \emph{length} of the matching window.
In sections~\ref{sec:WindowSpinAligned} and~\ref{sec:WindowPrecessing} below, we will describe our method of choosing the matching window, which is different for spin-aligned and precessing systems.

Recall that the hybridization method involves varying the PN parameters and frame parameters to minimize the PN-NR differences in the matching window. 
It is not appropriate to simply compare PN to NR in the window where the difference is optimized.
To better evaluate how well we match PN and NR, we also compare PN-NR differences in a new window, called a \emph{test window}, that is outside the matching window so that PN-NR differences are not minimized there, and that is earlier in time than the matching window so that both PN and NR should be valid there.  
The idea is that for a good choice of matching window, and for sufficiently accurate NR and PN waveforms, the PN-NR differences in the test window should also be small.
But it's still expected that the differences in the test window is larger than the differences in the matching window, given that both PN and NR (waveforms and parameters) are not accurate and disagree with each other, as we explained in Sec.~\ref{sec:PNparameter} and \ref{sec:Improvement}.
Unless otherwise specified, the test window is chosen to be 10 orbits long, and the center of the test window (in number of orbits) is 30 orbits earlier than the center of the matching window.

We also show the test window error for 10 long waveforms in Fig.~\ref{fig:ImprovementTest}. 
By comparing the results with and without optimization over PN parameters (brown and purple columns), we can see that for most systems, the optimization over PN parameters can result in smaller error in the test window.

\subsection{Spin-aligned systems}
\label{sec:WindowSpinAligned}
Figure \ref{fig:Window} shows how we determine the end time of the matching window for spin-aligned systems.
In this figure, we fix the length of the matching window as 15 orbits long, and vary the end time of the matching window.
We plot $\mathcal{E}$ in both the matching window and the test window as blue and green curves, respectively.
We can see that there is an optimal choice for the end time of the matching window, which is about $3000\,M$ to $8000\,M$ before merger, where the errors in the matching and test windows show a minimum.
For matching windows ending earlier than this optimum, the errors are larger because of larger eccentricity, as shown in Sec.~\ref{sec:error}.
The slope of the matching window error (blue curve) before $-5000\,M$ agrees very well with eccentricity damping, which is shown as the red dotted curve.
For matching windows ending later than the optimum, we get larger errors because PN breaks down at late times.

The other free parameter is the matching window's length. 
Generally speaking, the matching window should be as long as possible,
provided that the numerical error of the NR waveform is less than the systematic error stated in Sec.~\ref{sec:error}. 
However, most current NR simulations are short, and long NR simulations are very expensive. 
If the NR waveform is too short to include sufficient data, the hybridization model will be invalid.

\begin{figure}[tb]
\includegraphics[width=\onecol]{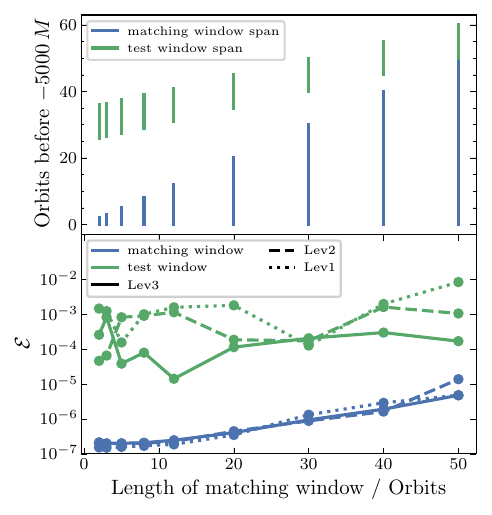}
\caption{The hybridization error for different lengths of matching window for spin-aligned system SXS:BBH:1412.
  We fix the end time of the matching window to be $5000\, M$ before merger. The test window is 10 orbits long, and the center of the test window (in number of orbits)
  is 30 orbits earlier than the center of the matching window.
  ``Lev'' labels the resolution of the NR simulations, as in
  Fig.~\ref{fig:SourceErrorPrecessing}.
}
\label{fig:WindowLength}
\end{figure}

\begin{figure}[tb]
\includegraphics[width=\onecol]{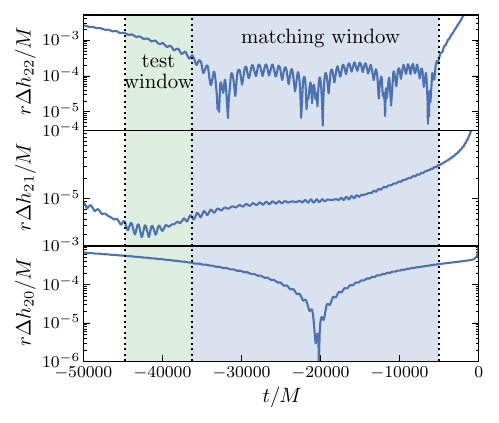}
\caption{The residual of spin-aligned system SXS:BBH:1412 corresponding to the rightmost points for Lev3 in Fig.~\ref{fig:WindowLength}. The matching window is 50 orbits long and ends at $5000\,M$ before merger, and the test window is 10 orbits long and immediately before the matching window. The error increases rapidly outside the matching window, so that the error in the test window is much larger than the error in the matching window. The error increases outside the matching window because of slight differences between the PN and NR waveforms. In this spin-aligned case, we use non-eccentric PN to hybridize with a slightly eccentric NR waveform. While we can minimize the error within the matching window, the inherent differences between the models prevent us from reducing the error outside.}
\label{fig:Error50Orbits}
\end{figure}

To find the minimal duration of the matching window, we fix the end time based
on Figure \ref{fig:Window}, and change the start time. 
We also use a test window to check the results.
The position of the matching window and test window are shown in the upper panel of Fig.~\ref{fig:WindowLength}, and the hybridization errors in both windows are shown in the lower panel.
The solid, dashed, and dotted curves in the lower panel show the errors using NR waveforms with different numerical resolution, labeled by ``Lev'' as in Section~\ref{sec:error} above.
We can see from the matching window errors (blue curves) that the NR numerical error does not significantly affect the hybridization error in the matching window. 
But when comparing the test window errors (green curves), we can see that they roughly converge to smaller values as we increase the resolution. This suggests that low-resolution NR simulations can accumulate numerical phase error over long periods of time. 
But note that the NR dephasing between the two highest Levs, $\Delta \phi$, does not directly reflect the inaccuracy of the NR waveforms, as the NR parameters within the matching window for different Levs can deviate by $10^{-2}$, given their evolution over nearly a hundred orbits prior to the matching window.
Consequently, the optimized PN parameters for various NR Levs also differ by $10^{-2}$, allowing the PN waveform to agree better with their respective NR waveforms and partially compensating for the phase error. Assessing the adequacy of the current NR resolution for hybridization purposes requires performing the entire hybridization process using various NR Levs, as shown in Fig.~\ref{fig:WindowLength}. However, the conclusion depends on the precision and frequency band needed for these observations (specifically, the frequency band of the test window). We leave this for future work.

We can see from the Lev3 errors (solid green and blue curves) that the optimal length of matching window should be around 5 to 20 orbits. 
The shorter the matching window is, the less information it includes, and the larger the chance that the small error in the matching window is spurious;
for very short matching windows ($<5$ orbits) the test window errors (green curves) no longer converge and the Lev3 test window error (solid green curve) increases significantly.
For longer matching windows ($>20$ orbits), the Lev3 test window error (green solid curve) becomes flat and the matching window errors (blue curves) increase. 
This is because on the one hand, the start time of the matching window is earlier, where the NR eccentricity is larger, resulting in a larger error in the matching window, as shown by the blue curve in Fig.~\ref{fig:Window}. 
On the other hand, a longer matching window leads to more accumulated phase error, potentially caused by both the disagreement between zero-eccentricity PN and small-but-finite-eccentricity NR, as well as insufficient NR resolution.
In Fig.~\ref{fig:Error50Orbits}, we plot the residual corresponding to the rightmost points for Lev3 in Fig.~\ref{fig:WindowLength}. 
We can see that even if the matching window is long enough to be adjacent to the test window, we still cannot achieve good alignment between PN and NR outside the matching window. 
The error increases beyond the matching window because the PN and NR waveforms are inherently not identical. 
In this case of spin-aligned systems, we use a non-eccentric PN model to hybridize with NR waveforms that possess slight eccentricity. 
Given the inherent differences between the waveform models, it is not possible to reduce the error outside the matching window, even if we achieve minimal error within it.

For NR waveforms with different and small but nonzero eccentricities, the best location of the matching window could be slightly different.
We can refer to Eq.~\eqref{eq:failure} and Eq.~\eqref{eq:EpsilonE}.
Error due to non-zero eccentricity and error from breakdown of PN are comparable when
\begin{equation}
   \text{Coefficient}\cdot e^2 \approx \left(\text{Coefficient}\cdot\Omega_\text{orb}^{(2N+1)/3}\right)^2
  \,.
\label{eq:failureSpinAligned}
\end{equation}
In Fig.~\ref{fig:Window}, the matching window error (blue curve) before $-8000\,M$ corresponds
to the left-hand side, the matching window error after $-3000\,M$ corresponds to the RHS,
and the region between $-8000\,M$ and $-3000\,M$ is where the two
effects are comparable.
For cases with different but small eccentricities, we expect the error
curve after $-3000\,M$ to be similar to that in Fig.~\ref{fig:Window}, and
the curve before $-8000\,M$ to be parallel to the blue curve in
Fig.~\ref{fig:Window}, but with a vertical shift.
Since the curve after $-3000\,M$ is steep and the curve before $-8000\,M$ is gradual, the point where the left and right part of the curve meet with each other (which is the best location for the matching window) is not very sensitive to the eccentricity of the system.

At the current state-of-the-art NR accuracy, and for $\mathcal{E}$ between PN and NR in the test window of $\sim 10^{-4}$, the minimal choice for a matching window for spin-aligned systems would be at least $5$ orbits long and ending $3000\,M$ before merger.  
This minimal choice will become longer and earlier if higher-accuracy hybrids are desired, and if the accuracy of the waveforms improves.
Matching windows longer than 20 orbits or ending earlier than $8000\,M$ are not beneficial given the current state of the art, but would be beneficial if the residual eccentricity of NR waveforms were decreased or if eccentric PN waveforms were used for hybridization and the potential degeneracy between eccentricity and supertranslations are properly handled.

\subsection{Precessing systems}
\label{sec:WindowPrecessing}

\begin{figure}[tb]
\includegraphics[width=\onecol]{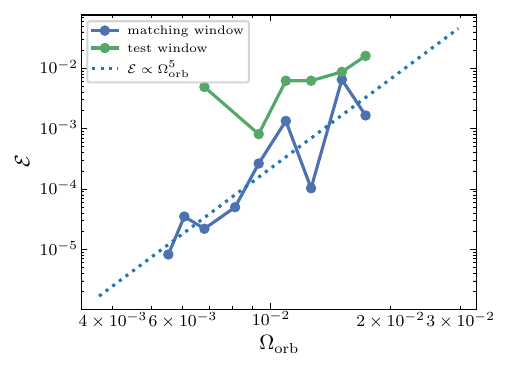}
\caption{The $L^2$ normalized error $\mathcal{E}$ in the matching window as a function of orbital angular velocity for system SXS:BBH:2621. The matching window is 10 orbits long, but the end time of the matching window is varied, and the orbital angular velocity is the mean value over the matching window. The test window is 10 orbits long, and the center of the test window (in number of orbits) is 30 orbits earlier than the center of the matching window.
}
\label{fig:WindowPre}
\end{figure}

In Sec.~\ref{sec:errorPrecessing}, we established that the error in hybridizing precessing systems mainly results from the absence of higher-order PN spin terms.
Because PN error decreases with smaller frequencies, we anticipate smaller hybridization errors at earlier times.
To determine the best location of the matching window, we plot the hybridization error of system SXS:BBH:2621 in the matching window versus orbital angular velocity $\Omega_{\text{orb}}$ in Fig.~\ref{fig:WindowPre}. 
This system has $\chi_p=0.49$.
Since the leading missing PN precession-related spin terms are expected to be $\propto (S_{\perp}/M^2)^\alpha v^\beta$, where $v=(\Omega_\text{orb}M)^{1/3}$ is the PN expansion parameter, we can get the indices $\alpha$ and $\beta$ by fitting the points in Fig.~\ref{fig:SourceErrorPrecessing} and Fig.~\ref{fig:WindowPre}, respectively.
Then we can get an empirical formula $\mathcal{E}\sim 3.6\times10^8\times(S_{\perp}/M^2)^{4.1}\Omega_\text{orb}^5 $, or
\begin{equation}
\Omega_\text{orb}\lesssim 0.012\mathcal{E}^{0.2}(S_{\perp}/M^2)^{-0.82}.
\label{eq:WindowPrecessing}
\end{equation}
Once a required precision $\mathcal{E}$ for a 10-orbit long window is given, we can use Eq.~\eqref{eq:WindowPrecessing} to determine the average orbital angular velocity of the matching window.  
From the average orbital angular velocity one can figure out the end time of the matching window.
Note that both Figs.~\ref{fig:SourceErrorPrecessing} and~\ref{fig:WindowPre} use a 10-orbit long matching window, because it is short enough to avoid being dominated by long-term accumulated phase error, and also limits significant frequency changes within the window.

\begin{figure}[tb]
\includegraphics[width=\onecol]{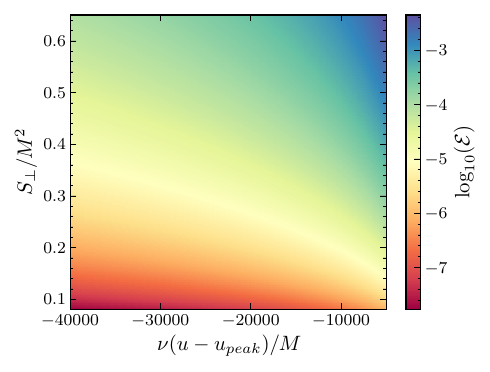}
\caption{The predicted $L^2$ normalized error $\mathcal{E}$ for a 10-orbit matching window based on Eq.~\ref{eq:WindowPrecessing}. The horizontal axis is the approximate center of the matching window, where $\nu=m_1m_2/(m_1+m_2)^2$, and $u$ is the retarded time. The vertical axis is the maximum of the in-plane component of the two dimensionless spin momenta. }
\label{fig:WindowPreChoice}
\end{figure}

We also plot the predicted $\mathcal{E}$ given a 10-orbit window in terms of the end time of the matching window and $S_{\perp}/M^2$ in Fig.~\ref{fig:WindowPreChoice}.
Here we used leading order PN to substitute orbital angular velocity with time: $\nu(u-u_{\mathrm{peak}})=\frac{5}{256}\Omega_{\text{orb}}^{-8/3}$, where $\nu=m_1m_2/(m_{1}+m_{2})^2$. 

Determining the ideal matching window length for precessing systems through experiments is challenging, unlike the spin-aligned cases. This challenge arises because the recommended end time for the matching window is significantly earlier for systems with large spin. Consequently, generating a complete precession cycle before this end time is impractical using NR simulations. However, considering that the error is primarily governed by the PN spin terms, which diminish at earlier times (in contrast to errors stemming from non-zero eccentricity, which worsen at earlier times), it is anticipated that longer matching windows are more beneficial for precessing systems.

\subsection{Short waveforms}
\label{sec:ShortWaveform}

\begin{figure}[tb]
\centering
\includegraphics[width=\onecol]{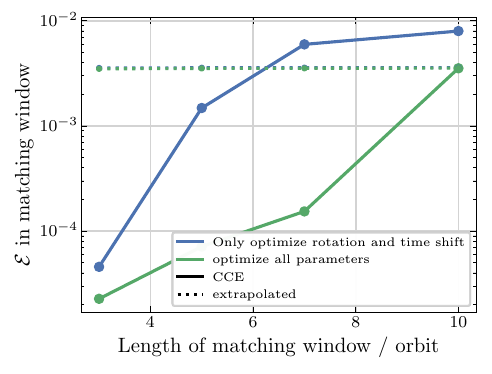}
\hspace{0 pt}
\includegraphics[width=\onecol]{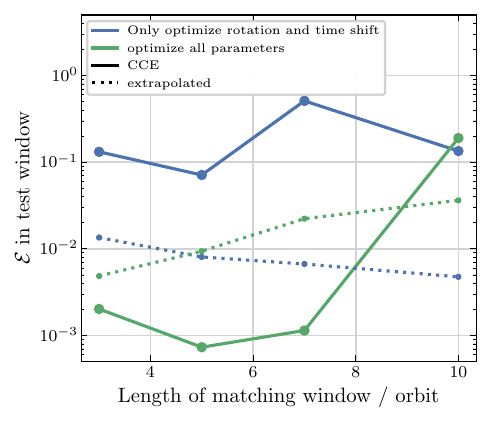}
\caption{The $L^2$ normalized error $\mathcal{E}$ for different lengths of the matching window for system SXS:BBH:2617. The test window is between 30 and 40 orbits before $-8000\,M$. We fix the start time of the matching window to be $8000\,M$ before merger. The upper and lower panel show $\mathcal{E}$ in the matching window and the test window, respectively. The dotted curves use extrapolated NR waveforms, while the solid curves use CCE NR waveforms. All the curves are obtained after adjusting the 3-D rotation and time shift, and all the curves using CCE waveforms are obtained after mapping to the PN BMS frame. In addition to the 3-D rotation and time shift, the green curves optimize all 12 parameters, which is the full hybridization procedure described in Section~\ref{sec:Procedure}.
}
\label{fig:WindowShort}
\end{figure}

\begin{figure}[tb]
\includegraphics[width=\onecol]{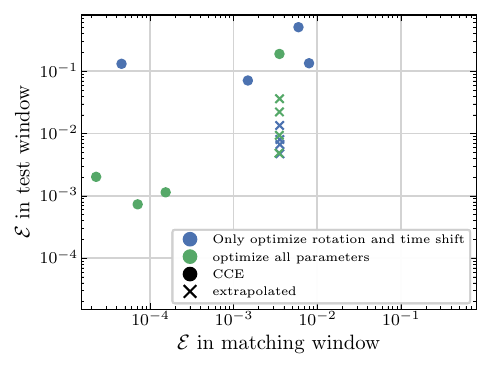}
\caption{The $L^2$ normalized error $\mathcal{E}$ for different lengths of the matching window for system SXS:BBH:2617. The test window is between 30 and 40 orbits before $-8000\,M$. We fix the start time of the matching window to be $8000\,M$ before merger. The crosses use extrapolated NR waveforms, while the dots use CCE NR waveforms. All the data are obtained after adjusting the 3-D rotation and time shift, and all the dots using CCE waveforms are obtained after mapping to the PN BMS frame. The blue and the green data points are the same data as the blue and green curves in Fig.~\ref{fig:WindowShort}.
The extrapolated waveforms all have the same error in the matching window, because for those waveforms the matching window error is dominated by incorrect gravitational-wave memory.
}
\label{fig:correlation}
\end{figure}

Unfortunately, most of the NR waveforms in current catalogs are not long enough to meet the minimal criteria for a matching window discussed in Sec.~\ref{sec:WindowChoice}.
In this scenario, it is important to note that the optimized PN parameters and BMS frame computed by the hybridization procedure may be biased by the breakdown of PN. 
Additionally, there is a risk of overfitting when optimizing the PN parameters and inertial frames from too-small a matching window.

To better understand these issues, we simulate an ``$8000\,M$'' waveform scenario. 
We set the start time of the matching window to be $-8000\,M$ and test the hybridization results using a 10-orbit long test window positioned between 30 and 40 orbits before $-8000\,M$.
The results are shown in Fig.~\ref{fig:WindowShort} and Fig.~\ref{fig:correlation}.

In Fig.~\ref{fig:WindowShort}, we display $\mathcal{E}$ for different lengths of the matching window for system SXS:BBH:2617 for this scenario.
The matching window errors obtained using extrapolated waveforms (dotted lines) in the upper panel are independent of the length of the matching window because the error is dominated by the incorrect gravitational-wave memory in the extrapolated waveforms. 
The matching window errors obtained using CCE waveforms (solid lines) show that as the length of the matching window becomes shorter, the error in the matching window becomes smaller. 
However, by comparing the errors in the matching window and test window (upper and lower panels), we can see that smaller error in the matching window does not mean smaller error in the test window. 
To more intuitively show the relation between the error in the matching window and the test window, we gather all the data points from Fig.~\ref{fig:WindowShort}, and plot them again in Fig.~\ref{fig:correlation}. 
We can see that the errors in the test window and matching window show no unambiguous trend.
This non-correlation of the matching and test window errors suggests that although we can always get a small error in the matching window if the matching window is short, it is not gaining us more information about the PN parameters and BMS frame to improve the error in the test window.

So, if we do not have NR waveforms that are long enough to meet the criteria outlined in Sec.~\ref{sec:WindowSpinAligned} and Sec.~\ref{sec:WindowPrecessing}, the hybridization could be biased by the breakdown of PN and overfitting. 

\section{Conclusions}
\label{sec:conclusion}
We provide a BBH waveform hybridization routine that includes every spin-weighted spherical harmonic mode with $-\ell \leq m \leq \ell$ for $\ell\leq 8$ as well as memory effects, and utilizes BMS frame fixing and PN parameter optimization.
The $L^2$ normalized error $\mathcal{E}$ between PN and NR waveforms in a 10-orbit \emph{matching} window is roughly $10^{-7}$ for spin-aligned systems, and roughly $10^{-5}$ for precessing systems.

However, a more informative measure of the accuracy of a hybridized waveform is the PN-NR difference in a \emph{test} window that is earlier than the matching window so that PN errors are small, and that does not overlap with the matching window so that we are not actively minimizing the PN-NR difference there.  
In a test window that is 10 orbits long and 30 orbits earlier than the matching window, we can reduce the normalized $L^2$ norm of PN-NR difference to less than $\sim 10^{-4}$ for spin-aligned systems, as shown by the solid green curve in Fig.~\ref{fig:WindowLength}.
This corresponds to $\sim 10^{-2}$ residual between PN and NR waveforms in this interval.
Again, we note that these numbers are not directly comparable to the frequency-domain mismatches used in GW data analysis, even assuming a flat noise curve, because the test and matching windows will not generally encompass the entire detectable waveform.  At best, these numbers constitute lower bounds for the full mismatch.  However, they do suggest that even for spin-aligned systems the results may be inadequate for high-SNR events in future detectors such as Cosmic Explorer and LISA.

For spin-aligned systems, the error in the test window can likely be decreased somewhat by using higher-accuracy NR waveforms, since the green curves in Fig.~\ref{fig:WindowLength} are roughly decreasing with NR resolution.  
However, the error is ultimately limited by trying to match zero-eccentricity PN waveforms with small but non-zero eccentricity NR waveforms. 
To achieve test window differences much better than $\mathcal{E} \lesssim
10^{-4}$ for spin-aligned systems, it will be necessary to either reduce the
residual eccentricity of the NR waveforms or to use an eccentric PN model for
hybridization.

While we recommend optimizing the intrinsic PN parameters in addition to the BMS frame parameters, we find that this optimization can sometimes lead to fractional differences in mass ratio and spin up to $\sim 10^{-1}$ between NR and the best-fit PN waveform (see Fig.~\ref{fig:Parameter}). 
This could arise from overfitting, where missing PN terms or NR truncation errors can be absorbed into the PN parameter optimization. 
This can be more significant for most current NR waveforms as they are typically much shorter than the ones considered here (see Sec.~\ref{sec:ShortWaveform} for tests using NR waveforms starting $8000M$ before merger). 
As PN continues to improve and NR waveforms become longer, we expect these differences to decrease. 
In the meantime, caution is warranted when varying the intrinsic PN parameters for short NR waveforms.

In principle, if one had perfect and exactly quasi-circular NR waveforms, the matching window should be chosen as early as possible (to avoid matching at late times near merger where PN becomes inaccurate), and it should be chosen as long as possible (so that one can match over a longer stretch of data and thus obtain a better fit).  
For current NR waveforms, and for spin-aligned systems, we find that to achieve $\mathcal{E}\sim 10^{-4}$ differences in the test window, the matching window should be at least $5$ orbits long and should end at least $3000\,M$ before merger.
While in principle a longer and earlier matching window should be better, we find that matching windows longer than 20 orbits or ending earlier than $8000\,M$ are not beneficial because the error is then dominated by the nonzero eccentricity of the NR waveforms as described above.
By using either smaller-eccentricity NR waveforms or employing
eccentric PN, it should be possible to further improve hybridization
by using longer and earlier matching windows.

Hybridization of precessing systems is much more difficult than spin-aligned systems, because the current PN waveforms for precessing systems are much less accurate than for spin-aligned systems. 
For precessing systems, the error in the hybrid is likely to be dominated by the lack of precession-related spin-spin terms in the PN model.
Unlike the error caused by non-zero eccentricity, the error stemming from PN terms diminishes at earlier times. So there is no optimal choice of the matching window even in the entire $10^5\,M$ span of the NR waveforms in the SXS catalog, and it always improves the hybridization by using matching windows as long and early as possible.
We provide a formula, Eq.~\ref{eq:WindowPrecessing}, and a plot, Fig.~\ref{fig:WindowPreChoice}, as a guide for choosing a matching window for precessing hybrids. 
The recommended end time for the matching window is significantly earlier for systems with large spin.

It is possible that using the effective one-body
(EOB) formalism \cite{Buonanno_1999}
instead of PN could significantly affect the position and length
of the matching window and decrease the requirement for long NR
waveforms.  However, to do so requires that one have $\Psi_M$ in the EOB
formalism to be able to fix the BMS freedoms, which is not
currently the case. Furthermore, current EOB waveforms are calibrated
using the extrapolated NR waveforms, which lack memory effects. We
plan to explore how the hybridization procedure changes when using
EOB waveforms trained on CCE waveforms in future work.

\begin{acknowledgments}

We thank Sizheng Ma, Nils Deppe, Qing Dai, Harald Pfeiffer and Aditya Vijaykumar for useful discussions. 
Computations for this work were preformed on the Wheeler cluster at Caltech, which is supported by the Sherman Fairchild Foundation and by Caltech, Resnick High Performance Computing (HPC) Cluster at the Caltech High Performance Computing Center, Frontera at the Texas Advanced Computing Center, and Urania HPC system at the Max Planck Computing and Data Facility. This work was supported in part by
the Sherman Fairchild Foundation, by NSF
Grants PHY-2207342 and OAC-2209655 at Cornell, and by
NSF Grants PHY-2309211, PHY-2309231, and OAC-2209656 at Caltech.
The work of L.C.S. was partially supported by NSF CAREER Award PHY-2047382 and a
Sloan Foundation Research Fellowship.
V.V.~acknowledges support from NSF Grant No. PHY-2309301; UMass Dartmouth’s
Marine and Undersea Technology (MUST) Research Program funded by the Office of
Naval Research (ONR) under Grant No. N00014-23-1–2141; and the European
Union’s Horizon 2020 research and innovation program under the Marie
Skłodowska-Curie grant agreement No.~896869.
\end{acknowledgments}

\def\bibsection{\section*{References}}
\bibliography{References}

\end{document}

%% file: figs/mapping.tex
\def \globalscale {6}
\begin{tikzpicture}[y=0.80pt, x=0.80pt, yscale=-\globalscale,
  xscale=\globalscale, inner sep=0pt, outer sep=0pt,
  basicstyle/.style={draw=black},
  coordAstyle/.style={draw=red},
  coordBstyle/.style={draw=blue,dashed},
  arrowstyle/.style={draw,thick,->,>={Stealth[round]}},]

  \begin{scope}[shift={(0,15)}]
  \path[basicstyle, path picture={
    \coordinate (Aright) at (path picture bounding box.east);
    \coordinate (Atop) at (path picture bounding box.north);
    \coordinate (Acenter) at (path picture bounding box.center);
    \coordinate (Abottom) at (path picture bounding box.south);
    \draw[step=4,coordAstyle]
    (path picture bounding box.north west) grid
    (path picture bounding box.south east);}
  ] (09.7924,00.8808) .. controls (08.5075,00.8683) and (08.1578,01.1058) ..
    (07.4064,01.3340) .. controls (06.4116,01.7815) and (05.5009,02.4161) ..
    (04.5457,02.9446) .. controls (03.6083,03.6348) and (02.7389,04.4410) ..
    (02.1042,05.4591) .. controls (00.9688,07.0359) and (00.6904,09.3396) ..
    (01.4962,11.1481) .. controls (02.1506,12.3886) and (03.2960,13.2890) ..
    (04.5266,13.7968) .. controls (06.3034,14.6274) and (08.2751,14.8292) ..
    (10.1892,14.6078) .. controls (12.0602,14.4133) and (13.8930,13.8668) ..
    (15.6478,13.1653) .. controls (16.9924,12.5110) and (18.3390,11.7554) ..
    (19.3130,10.5330) .. controls (20.0188,09.7622) and (20.4272,08.6776) ..
    (20.4619,07.6015) .. controls (20.4691,06.0412) and (19.6593,04.5888) ..
    (18.5631,03.6082) .. controls (17.7344,02.6221) and (11.4456,00.7920) ..
    (09.7924,00.8808) -- cycle;

    \node[inner sep=0pt] (CA) at ($(Abottom) + (0,1)$)
    {PN parameter space};

  \coordinate (circA) at ($(Acenter) + (3,1)$);

  \draw[coordAstyle,fill=red] (circA) circle (.3pt);
  \end{scope}


  \begin{scope}[shift={(0,0)}]
  \path[basicstyle, path picture={
    \coordinate (Bleft) at (path picture bounding box.west);
    \coordinate (Btop) at (path picture bounding box.north);
    \coordinate (Bcenter) at (path picture bounding box.center);
    \coordinate (Bbottom) at (path picture bounding box.south);
    \draw[step=4,coordBstyle]
    (path picture bounding box.north west) grid
    (path picture bounding box.south east);}
  ] (05.1502,01.3408) .. controls (03.8256,01.5213) and (02.3710,01.9365) ..
    (01.6116,03.1325) .. controls (00.2362,05.1152) and (00.5354,07.9139) ..
    (01.9723,09.7816) .. controls (03.5391,11.9410) and (06.1568,12.9830) ..
    (08.6963,13.4642) .. controls (10.4951,13.7821) and (12.4002,13.9161) ..
    (14.1434,13.2589) .. controls (16.1893,12.5273) and (18.1169,11.2429) ..
    (19.2285,09.3316) .. controls (19.8040,08.3477) and (20.3232,07.2673) ..
    (20.4425,06.1260) .. controls (20.3510,04.6501) and (19.6397,03.2660) ..
    (18.7569,02.1085) .. controls (17.9380,01.0721) and (16.6637,00.4528) ..
    (15.3424,00.4492) .. controls (13.4816,00.2613) and (11.5828,00.5633) ..
    (09.8900,01.3650) .. controls (08.1835,01.9110) and (06.3333,01.1324) ..
    (05.1502,01.3408) -- cycle;

  \node[inner sep=0pt] (CB) at ($(Btop) + (0,-1)$) {NR parameter space};

  \coordinate (circB) at ($(Bcenter) + (2,2)$);

  \draw[coordBstyle,solid,fill=blue] (circB) circle (.3pt);
  \end{scope}




  \begin{scope}[shift={(35,14)}]
  \pgfmathsetmacro{\cubex}{20}
  \pgfmathsetmacro{\cubey}{12}
  \pgfmathsetmacro{\cubez}{.4}

  \draw (0,0,0) -- ++(\cubex,0,0) -- ++(0,\cubey,0) -- ++(-\cubex,0,0) -- cycle;
  \draw (0,0,\cubez) -- ++(\cubex,0,0) -- ++(0,\cubey,0) -- ++(-\cubex,0,0) -- cycle;
  \draw (0,0,0) -- ++(0,0,\cubez);
  \draw (\cubex,0,0) -- ++(0,0,\cubez);
  \draw (0,\cubey,0) -- ++(0,0,\cubez);
  \draw (\cubex,\cubey,0) -- ++(0,0,\cubez);

  \coordinate (wavespacetop) at (0.5*\cubex,0,\cubez);
  \coordinate (wavespacebottom) at (0.5*\cubex,\cubey,0);


  \node[inner sep=0pt] (Wlabel2) at ($(wavespacebottom) + (0,2)$)
  {Waveform space ($\infty$-dim)};

  \begin{scope}[shift={(-5.5,-4)}]

  \path[coordAstyle] (00.5815,08.9449) .. controls (06.4970,06.7157) and (11.7012,05.3401) .. (15.6345,00.1081);
  \path[coordAstyle] (02.7145,11.8397) .. controls (09.5860,10.0967) and (15.5899,07.2824) .. (20.5708,01.9974);
  \path[coordAstyle] (05.3655,13.7290) .. controls (12.1430,11.7892) and (18.7063,08.6730) .. (23.9227,03.8562);
  \path[coordAstyle] (04.0553,07.6042) .. controls (06.3635,10.8851) and (06.7478,10.9288) .. (10.6554,11.9348);
  \path[coordAstyle] (09.2964,05.4712) .. controls (11.9867,07.7205) and (11.9581,08.6495) .. (16.2744,09.3106);
  \path[coordAstyle] (13.5319,02.4545) .. controls (16.7846,05.7381) and (17.0197,05.4665) .. (21.7287,05.7454);

  \path[coordBstyle] (00.0738,05.0875) .. controls (07.0873,05.7937) and (13.9405,03.3110) .. (20.4490,01.0528);
  \path[coordBstyle] (00.0635,05.0446) .. controls (01.3254,07.8936) and (03.1050,10.6847) .. (06.0359,12.1140);
  \path[coordBstyle] (05.9750,12.1140) .. controls (12.2815,11.7618) and (19.4077,10.6857) .. (25.0298,07.5957);
  \path[coordBstyle] (07.4681,04.8618) .. controls (08.1832,07.6797) and (10.1970,09.6655) .. (12.3910,11.4680);
  \path[coordBstyle] (13.1967,03.3991) .. controls (13.7346,06.3300) and (15.8776,08.7347) .. (18.3769,10.2247);
  \path[coordBstyle] (18.6511,01.7232) .. controls (19.4376,04.4009) and (20.3282,07.2946) .. (23.1000,08.5488);

  \end{scope}

  \coordinate (imcircA) at ($(wavespacetop) + (5,10)$);
  \draw[coordAstyle,fill=red] (imcircA) circle (.3pt);

  \coordinate (imcircB) at ($(imcircA) + (0.5,-2)$);
  \draw[coordBstyle,solid,fill=blue] (imcircB) circle (.3pt);

  \end{scope}

  \path[arrowstyle, draw=red]
  ($(circA) + (1,0) $)
  to[out=0, in=180]
  node [pos=0.4, below=1em, fill=white] {$h_{\text{PN}}$}
  ($(imcircA) + (-1,0)$);

  \path[arrowstyle, draw=blue]
  ($(circB) + (1,0) $)
  to[out=0, in=180]
  node [pos=0.4, above=1em, fill=white] {$h_{\text{NR}}$}
  ($(imcircB) + (-1,0)$);


  \node[inner sep=0pt] (waveB) at ($(wavespacetop) + (35,0)$) {\includegraphics[width=100pt]{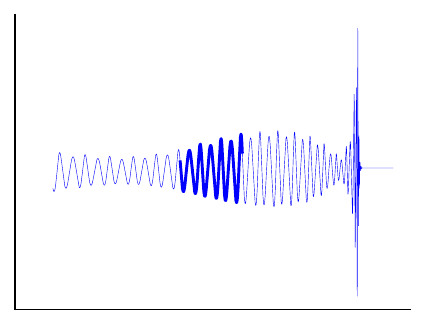}};
  \node[inner sep=0pt] (waveA) at ($(waveB) + (0, 18)$) {\includegraphics[width=100pt]{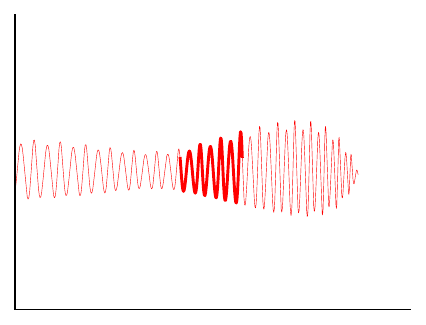}};

  \path[basicstyle] (imcircB) ++(0.5,0.5) -- (waveB.south west);
  \path[basicstyle] (imcircB) ++(0.5,-0.5) -- (waveB.north west);

  \path[basicstyle] (imcircA) ++(0.5,0.5) -- (waveA.south west);
  \path[basicstyle] (imcircA) ++(0.5,-0.5) -- (waveA.north west);

\end{tikzpicture}
